\documentclass[12pt]{article}

\usepackage{array}
\usepackage{epsfig}
\usepackage{amssymb}
\usepackage{graphics,graphpap}

\renewcommand{\thefootnote}{\fnsymbol{footnote}}

\newcommand{\ub}{\bar{u}}
\newcommand{\quark}{\langle \bar q q\rangle}
\newcommand{\mixed}{\langle \bar q \sigma gG q\rangle}
\newcommand{\squark}{\langle \bar s s\rangle}

\newcommand{\smixed}{\langle \bar s \sigma gG s\rangle}

\newcommand{\gluon}{\left\langle \frac{\alpha_s}{\pi}\,G^2\right\rangle}

\newcommand{\ds}{\displaystyle}

\def\s#1{\setbox0=\hbox{$#1$}%
  \rlap{\ifdim\wd0>.7em\kern.22\wd0\else\kern.1\wd0\fi /}#1}


\begin{document}

\begin{titlepage}
\begin{flushright}
\begin{tabular}{l}
IPPP/06/62\\
DCPT/06/124
\end{tabular}
\end{flushright}
\vskip1.5cm
\begin{center}
{\Large \bf \boldmath
Twist-3 Distribution Amplitudes of $K^*$ and $\phi$ Mesons}
\vskip1.3cm 
{\sc
Patricia Ball\footnote{Patricia.Ball@durham.ac.uk} and
G.W.~Jones\footnote{G.W.Jones@durham.ac.uk}}
  \vskip0.5cm
        {\em IPPP, Department of Physics,
University of Durham, Durham DH1 3LE, UK} 

\vskip2cm


\vskip5cm

{\large\bf Abstract\\[10pt]} \parbox[t]{\textwidth}{
We present a systematic study of twist-3 light-cone distribution 
amplitudes of $K^*$ and $\phi$ mesons in QCD. 
The structure of SU(3)-breaking corrections is studied in
detail. Non-perturbative input parameters are estimated from QCD sum
rules. As a by-product, we update the parameters describing the  
twist-3  distribution amplitudes of the $\rho$ meson. 
We also review and update predictions for
the twist-2 distribution amplitudes of $\rho$, $K^*$ and $\phi$.
}

\vfill

\end{center}
\end{titlepage}

\setcounter{footnote}{0}
\renewcommand{\thefootnote}{\arabic{footnote}}
\renewcommand{\theequation}{\arabic{section}.\arabic{equation}}

\newpage

\section{Introduction}
\setcounter{equation}{0}

In recent years  SU(3)-symmetry-breaking in processes involving light
vector mesons has attracted increasing interest. For some of these processes,
for instance the heavy-meson decays $B\to\rho\gamma$ vs.\ $B\to
K^*\gamma$, the uncertainty in SU(3) breaking is presently the
dominant source of theoretical error \cite{BVgamma,BJZ}. These and related
decays, like $B\to (\rho,K^*)\ell^+\ell^-$, are dominated, at
short distance, by flavour-changing neutral-current transitions which
are heavily suppressed in the Standard Model (SM) (they occur only at
loop-level) and hence are very sensitive to potential effects from new
physics. All these decays will be studied in detail at the LHC, with the
aim, in the case of the discovery of new physics in the TeV range, to 
elucidate its flavour structure, or,
in the case of continued absence of new particles in direct
searches, to constrain their possible masses and couplings. In any case, good
theoretical control over the SM predictions of such decays
is vital. The current theoretical approaches to describe them all
rely, in one way or the other, on their interpretation as hard 
exclusive reactions, with the hard scale
set by the heavy quark or meson mass which leads to an expansion in
terms of the inverse hard scale.\footnote{This also applies to
  lattice calculations, see Ref.~\cite{christalk}.}

In a perturbative framework, the method of choice for calculating
matrix elements of $B\to\,$light meson transitions is QCD
factorisation, which enters QCD sum rules on the light cone
\cite{LCSR}, QCD factorisation for non-leptonic and radiative $B$
decays \cite{QCDF} and perturbative QCD factorisation \cite{pQCD}. One
important ingredient in these calculations are light-cone 
hadron distribution amplitudes (DAs) which describe
the momentum-fraction distribution of partons at zero transverse separation 
in a particular Fock state, with a fixed number of constituents. DAs
are ordered by increasing twist; 
the leading-twist-2 meson DA $\phi_{2;M}$, which describes the momentum
 distribution of the valence quarks in the meson $M$, is related to the
 meson's Bethe--Salpeter wave function $\phi_{M,BS}$ by an integral
 over transverse momenta:
$$
\phi_{2;M}(u,\mu) = Z_2(\mu) \int^{|k_\perp| < \mu} \!\!d^2 k_\perp\,
\phi_{M,BS}(u,k_\perp).
$$
Here $u$ is the quark momentum fraction, $Z_2$ is the renormalisation
factor (in the light-cone gauge) for the quark-field operators in the
wave function, and $\mu$ denotes the renormalisation scale. For
pseudoscalar mesons, one has one twist-2 DA, whereas for vector mesons
there are two, $\phi_{2;M}^\perp$ and $\phi_{2;M}^\parallel$,
one for each independent polarisation state of the vector meson, transverse and
longitudinal, respectively. In this paper we study the twist-3
distribution amplitudes of the vector mesons $\rho$, $K^*$ and $\phi$,
with a particular emphasis on SU(3) (and G-parity) breaking
effects, and also update earlier results on twist-2 parameters. We do
not differentiate between $\rho$ and $\omega$ mesons as
their DAs only differ by the numerical values of
 hadronic parameters which, using the currently available
theoretical methods, coincide within errors.
Our paper is an extension of Ref.~\cite{BBL06} to vector-meson DAs
and finalises the preliminary results for twist-3 parameters quoted
earlier in Refs.~\cite{BVgamma,BJZ,frueher}. The results for twist-4
DAs will be published elsewhere.

The study of  vector-meson DAs has attracted
less attention than that of pseudoscalar DAs. The leading twist-2 DAs
of $\rho$ have been investigated in Ref.~\cite{BB96}, correcting a
mistake in the earlier literature \cite{CZreport}. 
The structure of twist-3 DAs of $\rho$, $K^*$ and $\phi$ and
their relation to twist-2 DAs have been studied in Ref.~\cite{BBKT},
but did not include all SU(3)-breaking effects. In this
paper, we complete the analysis of Ref.~\cite{BBKT} by including all
G-parity and SU(3) breaking corrections for $K^*$ and $\phi$ and also
providing numerical values of all hadronic parameters.

Our paper is organised as follows: in
Sec.~\ref{sec:2} we introduce notations and review the status of the
twist-2 parameters. In Sec.~\ref{sec:3} we provide parametrisations of
all twist-3 DAs to NLO in chiral expansion and including all G-parity
breaking contributions. In Sec.~\ref{sec:4} we discuss numerical
models for these DAs, based on the results from QCD sum rules. 
We conclude and summarise  in Sec.~\ref{sec:5}. The appendices contain
a discussion of the renormalisation-scale dependence of the twist-3
parameters, which is also affected by SU(3)-breaking corrections, and 
the QCD sum rules used to derive numerical values for all parameters.

\section{General Framework and Twist-2 DAs}\label{sec:2}
\setcounter{equation}{0}

In this section we introduce notations and review the status of
twist-2 parameters.

\subsection{Kinematics and Notations}
Light-cone meson DAs are defined in terms of matrix elements of
non-local light-ray operators extended along a certain light-like 
direction $z_\mu$, $z^2=0$, and sandwiched between the vacuum and the 
meson state.
We adopt the generic notation
\begin{equation}
\phi^\lambda_{t;M}(u),\ \psi^\lambda_{t;M}(u),\ \ldots
\end{equation}
and
\begin{equation}
\Phi^\lambda_{t;M}({\underline{\alpha}}),\ 
\Psi^\lambda_{t;M}({\underline{\alpha}}),\
\ldots
\end{equation}
for two-particle and  three-particle DAs, respectively. The
superscript $\lambda$ denotes the polarisation of the vector meson:
$\lambda=\parallel(\perp)$ for longitudinal (transverse)
polarisation. The first
subscript $t=2,3,4$ stands for the twist; the second one,
$M=\rho,K^*,\ldots$, specifies the meson. For definiteness, we will write
most expressions for $K^*$ mesons, i.e.\  $s\bar q$ bound states
with $q =u,d$. Whenever relevant, we will include quark mass
corrections in the form $m_s\pm m_q$, which allows one to obtain the
results for $\phi$ by $m_q\to m_s$. We do not include the $\omega$, as
all formulas for DAs coincide with those of the $\rho$; the difference
is in the numerical values of the hadronic parameters which, at least
in the framework of QCD sum rules, coincide with those for the $\rho$
except for small differences due to the difference in meson masses.
The variable $u$ in the definition of two-particle DAs
always refers to the momentum fraction carried by the quark, $u =
u_s$, whereas
$\bar u \equiv 1-u = u_{\bar q}$ is the antiquark momentum fraction.
The set of variables in the three-particle DAs,
$\underline{\alpha} = \{\alpha_1,\alpha_2,\alpha_3\} = \{\alpha_{s},
\alpha_{\bar q},\alpha_g\}$, corresponds to the momentum fractions 
carried by the quark, antiquark and gluon, respectively.

To facilitate the light-cone expansion, it is
convenient to use light-like vectors $p_\mu$ and $z_\mu$ instead of
the meson's 4-momentum $P_\mu$ and the coordinate $x_\mu$:
\begin{eqnarray}
  z_\mu &=& 
x_\mu-P_\mu\,\frac{1}{m_{K^*}^2}\left[xP -\sqrt{(xP)^2-x^2m^2_{K^*}}\,\right]
= x_\mu\left[1-\frac{x^2m_{{K^*}}^2}{4(zp)^2}\right]
-\frac{1}{2}p_\mu\,\frac{x^2}{zp}
    + \mbox{\cal O}(x^4)\,,
\nonumber\\
p_\mu &=& P_\mu-\frac{1}{2}\,z_\mu\, \frac{m^2_{K^*}}{pz}\,.
\label{lc-variables}
\end{eqnarray}
The meson's polarization vector 
$e^{(\lambda)}$ can be decomposed into projections 
onto the two light-like vectors and the orthogonal plane as follows:
\begin{eqnarray}
 e^{(\lambda)}_\mu &=& \frac{e^{(\lambda)}z}{pz}\, p_\mu +
                     \frac{e^{(\lambda)} p}{pz}\, z_\mu +
                     e^{(\lambda)}_{\perp\mu}
=  \frac{e^{(\lambda)} z}{pz}\left( p_\mu -\frac{m^2_{K^*}}{2pz}\, z_\mu
                                         \right)+e^{(\lambda)}_{\perp\mu}\,. 
\label{polv}
\end{eqnarray}
We also need the projector $g_{\mu\nu}^\perp$ 
onto the directions orthogonal to $p$ and $z$,
\begin{equation}
      g^\perp_{\mu\nu} = g_{\mu\nu} -\frac{1}{pz}(p_\mu z_\nu+ p_\nu z_\mu)\,, 
\end{equation}
and will often use the notations
\begin{equation}
    a_z\equiv a_\mu z^\mu, \qquad b_p\equiv b_\mu p^\mu
\end{equation}
for  arbitrary Lorentz vectors $a_\mu$ and $b_\mu$.

The dual gluon field strength
tensor is defined as $\widetilde{G}_{\mu\nu} =
\frac{1}{2}\epsilon_{\mu\nu \rho\sigma} G^{\rho\sigma}$. Our
convention for the covariant derivative is $D_\mu = \partial_\mu - i g
A_\mu$. Sometimes, a different convention for the sign of $g$ is used in the
literature, with $D_\mu = \partial_\mu + i g
A_\mu$. The sign of $g$ is relevant for all three-particle twist-3
DAs, see Tab.~\ref{tab:num}.

\subsection{Conformal Expansion and the 
Structure of SU(3)-Breaking Corrections}

A convenient tool to study DAs is provided by conformal expansion, see
Ref.~\cite{BKM03} for a review.\footnote{See Ref.~\cite{angi} 
for an alternative
approach not based on conformal expansion.}
The underlying idea is similar to partial-wave decomposition in 
quantum mechanics and allows one to separate
transverse and longitudinal variables in the Bethe--Salpeter 
wave--function.  The
dependence on transverse coordinates is formulated as scale dependence
of the relevant operators and is governed by
renormalisation-group equations, the dependence on the longitudinal
momentum fractions is described in terms of irreducible
representations of the corresponding symmetry group, the collinear
conformal group SL(2,$\mathbb R$). 

To construct the conformal expansion for an arbitrary multi-particle
distribution, one first has to decompose each constituent field into
components with fixed Lorentz-spin projection onto the
light-cone. Each such component has conformal spin
$$
j=\frac{1}{2}\, (l+s),
$$
where $l$ is the canonical dimension  and $s$ the (Lorentz-) spin
projection. In particular, $l=3/2$ for quarks and $l=2$ for gluons.
A  quark field is decomposed as $\psi_+ \equiv
\Lambda_+\psi$ and $\psi_-=
\Lambda_-\psi$ with spin projection operators $\Lambda_+ = 
\gamma_p\gamma_z/(2pz)$ and 
 $\Lambda_- = \gamma_z\gamma_p/(2pz)$, corresponding to
$s=+1/2$ and $s=-1/2$, respectively. For the gluon
field strength there are three possibilities:
$G_{z\perp}$ corresponds to $s=+1$,
$G_{p\perp}$ to $s=-1$, and both
$G_{\perp\perp}$ and $G_{zp}$ correspond to $s=0$.
Multi-particle states built of fields with definite Lorentz-spin
projection can be expanded in
irreducible  representations of SL(2,$\mathbb R$) 
with increasing conformal spin.
The explicit expression for the DA
of an $m$-particle state with the lowest possible conformal spin
 $j=j_1+\ldots+j_m$, the so-called asymptotic DA, is given by \cite{BF90}
\begin{equation}
\phi_{as}(\alpha_1,\alpha_2,\cdots,\alpha_m) =
\frac{\Gamma(2j_1+\cdots +2j_m)}{\Gamma(2j_1)\cdots \Gamma(2j_m)}\,
\alpha_1^{2j_1-1}\alpha_2^{2j_2-1}\ldots \alpha_m^{2j_m-1}.
\label{asym}
\end{equation}
Multi-particle irreducible representations with higher spin
$j+n,n=1,2,\ldots$,
are given by  polynomials of $m$ variables (with the constraint
$\sum_{k=1}^m \alpha_k=1$ ), which are orthogonal over
 the weight function (\ref{asym}). For the twist-2 and 3 two-particle
 DAs these are Gegenbauer polynomials, whereas  the twist-3
 three-particle DAs get expanded in Appell polynomials.

In this paper we are particularly interested in SU(3)-breaking
corrections to DAs. These corrections come from different sources:
\begin{itemize}
\item SU(3) breaking of hadronic parameters: these effects are
  partially known for twist-2 parameters, see
  Refs.~\cite{CZreport,BBKT,BB03}, but have not been studied for
  twist-3 parameters before;
\item G-parity breaking parameters: these are of parametric order 
$m_s-m_q$ and vanish in the
  limit of equal quark mass, i.e.\ for $\rho$ and $\phi$. For twist-2
  DAs, these have been calculated, 
  to lowest order in the conformal expansion in
  Refs.~\cite{CZreport,BBKT,BB03,BL04,BZ05}; they are unknown for
  twist-3 DAs;\footnote{The results given in
    Refs.~\cite{BVgamma,BJZ,frueher} were preliminary versions of those
    obtained in this paper.}
\item explicit quark mass corrections in $m_s\pm m_q$ to DAs and
  evolution equations: these affect only higher-twist DAs and are
  induced by the QCD equations of motion (EOM) which relate twist-3
  DAs to each other and to twist-2 DAs, see Sec.~\ref{sec:3}. The mass
  corrections to vector meson DAs
  have been calculated to twist-3 accuracy in Ref.~\cite{BBKT}; the
  effect on the evolution of DAs under a change of the renormalisation
  scale so far has only been investigated for
  pseudoscalar DAs \cite{BBL06}.  
\end{itemize}
We shall study all these effects in this paper.

Let us now see how these corrections affect twist-2 DAs.

\subsection{Twist-2 Distributions}\label{sec:2.3}

The twist-2 DAs $\phi_{2;K^*}^{\parallel,\perp}$ of $K^*$ mesons are 
defined in terms of the following
matrix elements of non-local operators ($\xi = 2u-1$) \cite{BBKT}:
\begin{eqnarray}
\lefteqn{\hspace*{-2cm}\langle 0 | \bar q(x) \gamma_\mu s(-x) |
 K^*(P,\lambda)\rangle = 
f_{K^*}^\parallel m_{K^*} \left\{ \frac{e^{(\lambda)}x}{Px}\,P_\mu
\int_0^1 du\, e^{i\xi Px} \left[ \phi_{2;K^*}^\parallel(u) +
  \frac{1}{4}\, m_{K^*}^2 x^2
  \phi^\parallel_{4;K^*}(u)\right]\right.}\nonumber\\
&&{} + \left( e^{(\lambda)}_\mu -
P_\mu\,\frac{e^{(\lambda)}x}{Px}\right) \int_0^1 du\,e^{i\xi
  Px}\,\phi_{3;K^*}^\perp(u)\nonumber\\
&&\left. - \frac{1}{2}\,x_\mu
\,\frac{e^{(\lambda)}x}{(Px)^2} \, m_{K^*}^2 \int_0^1 du\,e^{i\xi Px}\,
\left[ \psi_{4;K^*}^\parallel(u) + \phi_{2;K^*}^\parallel(u) - 2
  \phi_{3;K^*}^\perp(u)\right]\right\},\label{2.9}
\end{eqnarray}
\begin{eqnarray}
\lefteqn{
\langle 0 | \bar q(x) \sigma_{\mu\nu} s(-x) | K^*(P,\lambda)\rangle 
=}\hspace*{1cm}
\nonumber\\
&&{}i f_{K^*}^\perp \left\{ (e_\mu^{(\lambda)} P_\nu - e_\nu^{(\lambda)}
  P_\mu) \int_0^1 du \, e^{i\xi Px} \left[ \phi_{2;K^*}^\perp(u) +
  \frac{1}{4}\, m_{K^*}^2 x^2
  \phi^\perp_{4;K^*}(u)\right]\right.
\nonumber\\
&&{}
+ (P_\mu x_\nu - P_\nu
x_\mu)\,\frac{e^{(\lambda)}x}{(Px)^2}\,m_{K^*}^2 \int_0^1 du \,
e^{i\xi Px} \left[ \phi_{3;K^*}^\parallel(u) -
  \frac{1}{2}\phi_{2;K^*}^\perp(u) -
  \frac{1}{2}\psi_{4;K^*}^\perp(u)\right]
\nonumber\\
&&{}\left.
+ \frac{1}{2}\,(e_\mu^{(\lambda)} x_\nu - e_\nu^{(\lambda)}x_\mu)\,
\frac{m_{K^*}^2}{Px} \int_0^1 du \,
e^{i\xi Px} \left[ \psi_{4;K^*}^\perp(u)-\phi_{2;K^*}^\perp(u)
  \right]\right\}.\label{2.10}
\end{eqnarray}
All other DAs in the above relations are of twist 3 or 4. We have
neglected all terms in the light-cone expansion which are of twist
5 or higher.
The normalisation of all DAs is given by
\begin{equation}
\int_0^1 du\, \phi(u) = 1\,.
\end{equation}
The above DAs are related to those defined in Refs.~\cite{BBKT,BB98} by
\begin{equation}
\addtolength{\arraycolsep}{5pt}
\begin{array}[b]{r@{\ =\ }l@{\qquad}r@{\ =\ }l@{\qquad}r@{\ =\ }l}
\phi_{2;K^*}^{\parallel(\perp)} & \phi_{\parallel(\perp)}\,, &
\phi_{3;K^*}^\parallel & h_\parallel^{(t)}\,, & \psi_{4;K^*}^\parallel &
g_3\,,\\[10pt]
\phi_{4;K^*}^{\parallel(\perp)} & {\mathbb A}_{(T)}\,, &
\phi_{3;K^*}^\perp & g_\perp^{(v)}\,, & \psi_{4;K^*}^\perp &
h_3\,.
\end{array}
\addtolength{\arraycolsep}{-5pt}
\end{equation}

The conformal expansion of $\phi_2^{\parallel,\perp}$ reads
\begin{equation}\label{eq:confexp}
\phi_2^{\parallel,\perp}(u) = 6 u \bar u \left\{ 1 + \sum_{n=1}^\infty
a_n^{\parallel,\perp} C_n^{3/2}(2u-1)\right\} 
\end{equation}
in terms of the (non-perturbative) Gegenbauer moments
$a_n^{\parallel,\perp}$ and the
Gegenbauer polynomials $C_n^{3/2}$.
To leading-logarithmic accuracy, the  
$a_n$ renormalise multiplicatively as
\begin{equation}
a^{\rm LO}_n(\mu^2) = L^{\gamma^{(0)}_n/(2\beta_0)}\, a_n(\mu_0^2),
\end{equation}
where $L = \alpha_s(\mu^2)/\alpha_s(\mu_0^2)$,
$\beta_0=(33-2N_f)/3$, and
the anomalous dimensions $\gamma^{(0)}_n$ are given by \cite{andim1}
\begin{eqnarray*}
\gamma^{\parallel(0)}_n &=&  8C_F \left(\psi(n+2) + \gamma_E - \frac{3}{4} -
  \frac{1}{2(n+1)(n+2)} \right),\\
\gamma^{\perp(0)}_n &=&  8C_F \left(\psi(n+2) + \gamma_E -
  \frac{3}{4} \right).
\end{eqnarray*}
To next-to-leading order accuracy, the scale dependence of the
Gegenbauer moments is more complicated and reads 
\cite{Mueller}
\begin{equation}
 a^{\rm NLO}_n(\mu^2) =  a_n(\mu_0^2) E_n^{\rm NLO} 
+\frac{\alpha_s}{4\pi}\sum_{k=0}^{n-2} a_k(\mu_0^2)\,  
L^{\gamma_k^{(0)}/(2\beta_0)}\, d^{(1)}_{nk},  
\end{equation} 
where 
$$
E_n^{\rm NLO} =  L^{\gamma^{(0)}_n/(2\beta_0)}\left\{1+ 
            \frac{\gamma^{(1)}_n \beta_0 -\gamma_n^{(0)}\beta_1}{8\pi\beta_0^2}
                 \Big[\alpha_s(\mu^2)-\alpha_s(\mu_0^2)\Big]\right\}
$$
with $\beta_1 = 102-(38/3)N_f$; 
$\gamma^{(1)}_n$ are the diagonal two-loop anomalous dimensions, which
have been calculated, for the vector current, in Ref.~\cite{Floratos},
and, for the tensor current, in Ref.~\cite{Haya}.
The mixing coefficients $d^{(1)}_{nk}$, $k\le n-2$, 
are given, in closed form in Ref.~\cite{Mueller}, for the axial
vector current; the formulas are valid for
arbitrary currents upon substitution of the corresponding one-loop
anomalous dimension.

For the lowest moments $n=0,1,2$ one has, explicitly:
$$
\gamma_0^{\parallel(1)} =0\,, \qquad 
\gamma_1^{\parallel(1)} = \frac{23110}{243} - \frac{512}{81}\, N_f\,,
   \qquad  
\gamma_2^{\parallel(1)} =  \frac{34072}{243}-\frac{830}{81}\, N_f\,,
$$
\begin{equation}
\gamma_0^{\perp(1)} =\frac{724}{9} - \frac{104}{27}\,N_f\,, \qquad 
\gamma_1^{\perp(1)} = 124 - 8 N_f\,,\qquad  
\gamma_2^{\perp(1)} =  \frac{38044}{243}-\frac{904}{81}\, N_f\,,
\end{equation}  
and
\begin{eqnarray}
  d^{\parallel(1)}_{20} & = &
  \frac{35}{9}\,\frac{20-3\beta_0}{50-9\beta_0}
  \left(1-L^{50/(9\beta_0)-1}\right), \nonumber\\
  d^{\perp(1)}_{20} & = &
  \frac{28}{9}\,\frac{16-3\beta_0}{40-9\beta_0}
  \left(1-L^{40/(9\beta_0)-1}\right).
\end{eqnarray}

Let us now review the numerical values of the twist-2 parameters, to
NLO in conformal spin. The longitudinal decay constants of the charged
mesons $\rho^\pm$, $K^{*\pm}$ can be extracted from the branching
ratios of $\tau^-\to V^- \nu_\tau$, whereas
$f^\parallel_{\rho,\omega,\phi}$ follow from $e^+e^-\to V^0$. A
critical discussion of the results, including effects of
$\rho$--$\omega$ and $\omega$--$\phi$ mixing, was given in
Ref.~\cite{BJZ} from which we quote the following results:
\begin{equation}
f_\rho^\parallel = (216\pm 3)\,{\rm MeV},\quad f_\omega^\parallel =
(187\pm 5)\, {\rm MeV},\quad f_{K^*}^\parallel = (220\pm 5)\,{\rm
  MeV},\quad f_\phi^\parallel = (215\pm 5)\,{\rm MeV}\,.
\end{equation}

The transverse decay constants $f^\perp$, on the other hand, cannot be
determined from experiment, but have to be calculated using
non-perturbative methods. Currently available results include QCD sum
rule determinations \cite{BB96,BBKT,BB03,BZ05} and lattice
calculations \cite{lattperp}. The corresponding results have been critically
reviewed and averaged in Ref.~\cite{BJZ}, with the following results:
\begin{equation}
f_\rho^\perp = (165\pm 9)\,{\rm MeV},\quad f_\omega^\perp =
(151\pm 9)\, {\rm MeV},\quad f_{K^*}^\perp = (185\pm 10)\,{\rm
  MeV},\quad f_\phi^\perp = (186\pm 9)\,{\rm MeV}\,.
\end{equation}

Let us now turn to the Gegenbauer moments
  $a_{1,2}^{\parallel,\perp}$. At present, there are no lattice
  determinations for any of those, so all available determinations
  come from QCD sum rules \cite{BBKT,BB03,BL04,BZ05,Bakulev} or quark models
  \cite{choi}. For mesons with definite G parity
  (equal mass quarks), i.e.\ $\rho$ and $\phi$ in our case,
  $a_1^{\parallel,\perp} =0$. For $a_1^{\parallel,\perp}(K^*)$, the
  results from QCD sum rule calculations converge to \cite{BZ05}
\begin{eqnarray}
a_1^\parallel(K^*)^{\mu=1\,{\rm GeV}} & = & 0.03 \pm 0.02\,,\qquad
a_1^\parallel(K^*)^{\mu=2\,{\rm GeV}}  = 0.02\pm 0.02 ,\qquad\nonumber\\
a_1^\perp(K^*)^{\mu=1\,{\rm GeV}} & = & 0.04 \pm 0.03\,,\qquad
a_1^\perp(K^*)^{\mu=2\,{\rm GeV}}  =  0.03\pm 0.03\,.
\end{eqnarray}

As for $a_2$, $a_2^{\perp,\parallel}(\rho)$ have been determined in
Ref.~\cite{BB96} and reinvestigated,recently, in
Ref.~\cite{BVgamma}, using the updated hadronic input collected in 
Tab.~\ref{tab:QCDSRinput}. The resulting values
\begin{eqnarray}
a_2^\parallel(\rho)^{\mu=1\,{\rm GeV}} & = & 0.15 \pm 0.07\,,\qquad
a_2^\parallel(\rho)^{\mu=2\,{\rm GeV}}  = 0.10\pm 0.05\,,\qquad\nonumber\\
a_2^\perp(\rho)^{\mu=1\,{\rm GeV}} & = & 0.14 \pm 0.06\,,\qquad
a_2^\perp(\rho)^{\mu=2\,{\rm GeV}}  =  0.11\pm 0.05\,,
\end{eqnarray}
are slightly smaller than those quoted in Ref.~\cite{BB96}. 
The value of $a_2^{\parallel,\perp}(K^*)$ has been determined in
Ref.~\cite{BBKT,BB03} and reinvestigated in Ref.~\cite{BVgamma}. 
The result is
\begin{eqnarray}
a_2^\parallel(K^*)^{\mu=1\,{\rm GeV}} & = & 0.11 \pm 0.09\,,\qquad
a_2^\parallel(K^*)^{\mu=2\,{\rm GeV}}  = 0.08\pm 0.06\,,\qquad\nonumber\\
a_2^\perp(K^*)^{\mu=1\,{\rm GeV}} & = & 0.10 \pm 0.08\,,\qquad
a_2^\perp(K^*)^{\mu=2\,{\rm GeV}}  =  0.08\pm 0.06\,.
\end{eqnarray}
The corresponding parameters of the  $\phi$ have received far less attention:
Ref.~\cite{BBKT} quotes $a_2^{\parallel,\perp}(\phi)^{\mu=1\,{\rm GeV}} 
= 0\pm 0.1$ and  Ref.~\cite{BJZ} $a_2^\perp(\phi)^{\mu=1\,{\rm GeV}}=0.2\pm
0.2$. In this paper, we evaluate the sum rules collected in
App.~\ref{app:A}, which include all relevant corrections in $m_s^2$
and in particular the radiative corrections to the quark condensate
term in $m_s\langle\bar s s\rangle$, and find
\begin{eqnarray}
a_2^\parallel(\phi)^{\mu=1\,{\rm GeV}} & = & 0.18 \pm 0.08\,,\qquad
a_2^\parallel(\phi)^{\mu=2\,{\rm GeV}}  = 0.13\pm 0.06\, ,\qquad\nonumber\\
a_2^\perp(\phi)^{\mu=1\,{\rm GeV}} & = & 0.14 \pm 0.07\,,\qquad
a_2^\perp(\phi)^{\mu=2\,{\rm GeV}}  =  0.11 \pm 0.05\,.
\end{eqnarray}

In summary, it is probably fair to say that all known determinations
of $a_{1,2}^{\perp,\parallel}$ point at fairly small values at
$1\,$GeV and that within the present accuracy
$a^\perp_{1(2)}=a^\parallel_{1(2)}$. 

Let us now turn to twist-3 DAs.

\section{Twist-3 Distributions}\label{sec:3}
\setcounter{equation}{0}

To twist-3 accuracy, there is a total of four two-particle DAs and three
three-particle DAs whose mutual interrelations have been unravelled in
Ref.~\cite{BBKT}, including quark mass corrections. 
The crucial point in constructing higher-twist DAs is the 
necessity to satisfy the QCD EOM which yield relations between 
physical effects of different origin: for example, using EOM, 
the contributions of orbital angular momentum in the 
valence component of the wave function can be expressed
in terms of contributions of higher 
Fock states. An appropriate framework for implementing these
constraints was developed in Ref.~\cite{BF90}: it is based on the derivation 
of EOM relations for non-local light-ray operators~\cite{string},
which are solved order by order in 
the conformal expansion; see Ref.~\cite{BKM03} for a review and
further references. In this way one can construct
self-consistent approximations for the DAs, which involve
a minimum number of hadronic parameters. The EOM
relations relating twist-2 and -3 DAs of vector mesons were derived in
Ref.~\cite{BBKT}, including all quark-mass corrections.  
What is new in the present paper is the inclusion of G-parity breaking 
corrections to three-particle DAs, which, via the EOM relations, also 
impact on the two-particle DAs. Based on these relations, we derive,
in this section, complete formulas for all
twist-3 DAs to NLO in the conformal expansion, including all G-parity
breaking effects. 
A non-zero quark mass also induces a mixing of
twist-2 parameters into those of twist-3 under a change of the
renormalisation scale. We also derive the corresponding
scaling relations. 

Let us start by defining the relevant DAs. The two-particle twist-3 DAs
$\phi_{3;K^*}^{\perp,\parallel}$ have already been defined in
Eqs.~(\ref{2.9}) and (\ref{2.10}). There are two more two-particle DAs,
$\psi_{3;K^*}^{\perp,\parallel}$, defined as%
\footnote{In the notations of Ref.~\cite{BBKT}, $\psi_{3;K^*}^\perp =
  \{1-(f_{K^*}^\perp/f_{K^*}^\parallel) (m_s+m_q)/m_{K^*}\}
g_\perp^{(a)}$, $\psi_{3;K^*}^\parallel =
  \{1-(f_{K^*}^\parallel/f_{K^*}^\perp) (m_s+m_q)/m_{K^*}\}
 h_\parallel^{(s)}$.}
\begin{eqnarray}
\langle 0|\bar q(z) \gamma_{\mu} \gamma_{5}
s(-z)|K^*(P,\lambda)\rangle
&=& \frac{1}{2}\,f_{K^*}^\parallel
m_{K^*} \epsilon_{\mu}^{\phantom{\mu}\nu \alpha \beta}
e^{(\lambda)}_{\nu} p_{\alpha} z_{\beta}
\int_{0}^{1} \!du\, e^{i \xi p  x} \psi_{3;K^*}^\perp(u)\,,\\
\langle 0|\bar q(z)s(-z)|K^*(P,\lambda)\rangle
 & = & {} -i f_{K^*}^\perp(e^{(\lambda)} z) m_{K^*}^{2}
\int_{0}^{1} \!du\, e^{i \xi p  z} \psi_{3;K^*}^\parallel(u)\,.
\end{eqnarray}
The normalisation is given by
\begin{equation}
\int_0^1 du\, \psi_{3;K^*}^{\parallel(\perp)}(u) = 1-
\frac{f_{K^*}^{\parallel(\perp)}}{f_{K^*}^{\perp(\parallel)}}\,
\frac{m_s+m_q}{m_{K^*}}\,,
\end{equation}
which differs from Ref.~\cite{BBKT}, where all DAs
were normalised to 1. The reason is that in \cite{BBKT} we
implicitly expanded the normalisation factor
$1/\{1-(f_{K^*}^{\parallel(\perp)}/f_{K^*}^{\perp(\parallel)})
(m_s+m_q)/m_{K^*}\}$ in powers of $m_s+m_q$, whereas in this paper we
 keep the full dependence on the quark masses.

There are also three three-particle DAs of twist 3:
\begin{eqnarray}
\langle 0 | 
\bar q(z) g \widetilde{G}_{\beta z}(vz)\gamma_z\gamma_5 s(-z) |
K^*(P,\lambda)\rangle & = & f_{K^*}^\parallel m_{K^*} (pz)^2
e^{(\lambda)}_{\perp\beta}
\widetilde\Phi_{3;K}^\parallel(v,pz)+\dots\,,
\nonumber\\
\langle 0 | 
\bar q(z) g G_{\beta z}(vz)i\gamma_z s(-z) | K^*(P,\lambda)\rangle & = & 
f_{K^*}^\parallel m_{K^*} (pz)^2
e^{(\lambda)}_{\perp\beta} \Phi_{3;K}^\parallel(v,pz)+\dots\,,\nonumber\\
\langle 0 | 
\bar q(z) g G_{z\beta}(vz)\sigma_{z\beta} s(-z) |
K^*(P,\lambda)\rangle & = & f_{K^*}^\perp m_{K^*}^2
(e^{(\lambda)}z)(pz) \Phi_{3;K^*}^\perp(v,pz)\,,\label{3.4}
\end{eqnarray}
where the dots denote terms of higher twist and we use the short-hand notation 
\begin{equation}
{\cal F}(v,pz) = \int {\cal
  D}\underline{\alpha}\,e^{-ipz(\alpha_2-\alpha_1+v\alpha_3)}
  {\cal F}(\underline{\alpha}) 
\end{equation}
with ${\cal F}(\underline{\alpha})$ being a three-particle DA.
$\underline{\alpha}$ is the set of parton momentum fractions
$\underline{\alpha} = \{\alpha_1,\alpha_2,\alpha_3\}$ and the integration
measure ${\cal D}\underline{\alpha}$ is defined as
\begin{equation}\label{eq:measure}
\int{\cal D}\underline{\alpha} \equiv \int_0^1 d\alpha_1 d\alpha_2
d\alpha_3\, \delta(1-\sum \alpha_i)\,.
\end{equation}

As discussed in Ref.~\cite{BBKT}, all these DAs are interconnected by
the QCD EOM. The analysis of these EOM and the resulting
relations including quark mass corrections in $m_s\pm m_q$ 
is the subject of Ref.~\cite{BBKT}, so we do not
repeat it here, but just quote the results:
\begin{eqnarray}
\psi_{3;K^*}^\parallel(u) & = & \ub\int_0^u dv\,\frac{1}{\bar v}\,
\Upsilon(v) + u \int_u^1 dv\,\frac{1}{v}\, \Upsilon(v)\,,\nonumber\\
\phi_{3;K^*}^\parallel(u) 
& = & \frac{1}{2}\,\xi\left[\int_0^u dv\,\frac{1}{\bar v}\,
\Upsilon(v) - \int_u^1 dv\,\frac{1}{v}\, \Upsilon(v)\right] + 
\frac{f_{K^*}^\parallel}{f_{K^*}^\perp}\,\frac{m_s+m_q}{m_{K^*}}\,
\phi_{2;K^*}^\parallel(u)\nonumber\\
&& {}+\frac{d}{du}\,\int_0^u d\alpha_1\int_0^{\bar u} d\alpha_2
\frac{1}{\alpha_3}
\,{\Phi}^\perp_{3;K^*}(\underline{\alpha})
\end{eqnarray}
with
\begin{eqnarray}
\Upsilon(u) & = & 2\phi_{2;K^*}^\perp(u) - 
\frac{f_{K^*}^\parallel}{f_{K^*}^\perp}\,\frac{m_s+m_q}{m_{K^*}} \left[
1 - \frac{1}{2}\,
\xi\frac{d}{du}\right]\phi_{2;K^*}^\parallel(u) -
\frac{1}{2}\,\frac{f_{K^*}^\parallel}{f_{K^*}^\perp}\,
\frac{m_s-m_q}{m_{K^*}}\,\frac{d}{du}\,\phi_{2;K^*}^\parallel(u)\nonumber\\
& & {}+\frac{d}{du}\,\int_0^u\!\!d\alpha_1\int_0^{\bar
u}\!\! d\alpha_2
\,\frac{1}{\alpha_3}\left(\alpha_1\,\frac{d}{d\alpha_1} +
\alpha_2\,\frac{d}{d\alpha_2}\, - 1\right) 
\Phi_{3;K^*}^\perp(\underline{\alpha})
\end{eqnarray}
and
\begin{eqnarray}
\psi_{3;K^*}^\perp(u)& = & \ub\int_0^u dv\,\frac{1}{\bar v}\,
  \Omega(v) + u \int_u^1 dv\,\frac{1}{v}\, \Omega(v)\,,\nonumber\\
\phi_{3;K^*}^\perp(u) 
& = & \frac{1}{4}\left[\int_0^u dv\,\frac{1}{\bar v}\,
\Omega(v) + \int_u^1 dv\,\frac{1}{v}\, \Omega(v)\right] + 
\frac{f_{K^*}^\perp}{f_{K^*}^\parallel}\,\frac{m_s+m_q}{m_{K^*}}\,
\phi_{2;K^*}^\perp(u)\nonumber\\
&& {}+\frac{d}{du}\,\int_0^u d\alpha_1\int_0^{\bar u} d\alpha_2
\frac{1}{\alpha_3}
\,{\Phi}^\parallel_{3;K^*}(\underline{\alpha})\nonumber\\
&&{} + \int_0^u d\alpha_1\int_0^{\bar u}
d\alpha_2\,\frac{1}{\alpha_3}\left( \frac{d}{d\alpha_1} +
\frac{d}{d\alpha_2} \right) 
\widetilde{\Phi}_{3;K^*}^\parallel(\underline{\alpha})
\end{eqnarray}
with
\begin{eqnarray}
\Omega(u)  &=&  2\phi_{2;K^*}^\parallel(u) + 
\frac{f_{K^*}^\perp}{f_{K^*}^\parallel}\,\frac{m_s+m_q}{m_{K^*}}\, 
\xi\,\frac{d}{du}\,\phi_{2;K^*}^\perp(u)-
\frac{f_{K^*}^\perp}{f_{K^*}^\parallel}\,\frac{m_s-m_q}{m_{K^*}}\, 
\frac{d}{du}\,\phi_{2;K^*}^\perp(u)\nonumber\\
& & {}+2\frac{d}{du}\,\int_0^u\!\!d\alpha_1
\int_0^{\bar u}\!\! d\alpha_2
\,\frac{1}{\alpha_3}\left(\alpha_1\,\frac{d}{d\alpha_1} +
\alpha_2\,\frac{d}{d\alpha_2}\right)
\Phi_{3;K^*}^\parallel(\underline{\alpha})\nonumber\\
&&{} +2\,\frac{d}{du}\,\int_0^u\!\!d\alpha_1
\int_0^{\bar u}\!\! d\alpha_2
\,\frac{1}{\alpha_3}\left(\alpha_1\,\frac{d}{d\alpha_1} -
\alpha_2\,\frac{d}{d\alpha_2}\right) 
\widetilde\Phi_{3;K^*}^\parallel(\underline{\alpha})\,.
\end{eqnarray}

The twist-3 three-particle DAs correspond to the light-cone projection
$\gamma_z G_{z\perp}$ and $\sigma_{\perp z} G_{\perp z}$,
respectively, which picks up the $s=\frac{1}{2}$ component of the
quark fields and the $s=1$ component of the gluonic field strength
tensor. According to (\ref{asym}), the (normalised) 
asymptotic DA is then given by $360\alpha_1\alpha_2\alpha_3^2$.
To NLO in the conformal expansion, each three-particle twist-3 DA involves
three hadronic parameters, which we label in the following way:
$\zeta,\kappa$ are LO and $\omega,\lambda$ NLO parameters. 
$\zeta$ and $\omega$ are G-parity conserving, whereas 
$\kappa$ and $\lambda$ violate G-parity and hence vanish for mesons
with quarks of equal mass, i.e.\  $\rho$ and $\phi$. We then have
\begin{eqnarray}
\Phi_{3;K^*}^\parallel(\underline{\alpha})
 & = & 360\alpha_1\alpha_2\alpha_3^2 \left\{
\kappa_{3K^*}^\parallel + \omega_{3K^*}^\parallel (\alpha_1-\alpha_2) +
\lambda_{3K^*}^\parallel \frac{1}{2}\,(7\alpha_3 -
3)\right\},\nonumber\\
\widetilde\Phi_{3;K^*}^\parallel(\underline{\alpha}) & = & 
360\alpha_1\alpha_2\alpha_3^2 \left\{
\zeta_{3K^*}^\parallel + \widetilde\lambda_{3K^*}^\parallel (\alpha_1-\alpha_2) +
\widetilde\omega_{3K^*}^\parallel \frac{1}{2}\,(7\alpha_3 -
3)\right\},\nonumber\\
\Phi_{3;K^*}^\perp(\underline{\alpha}) 
& = & 360\alpha_1\alpha_2\alpha_3^2 \left\{
\kappa_{3K^*}^\perp + \omega_{3K^*}^\perp (\alpha_1-\alpha_2) +
\lambda_{3K^*}^\perp \frac{1}{2}\,(7\alpha_3 -
3)\right\}.\label{3.11}
\end{eqnarray}
The relation to the parameters used in Ref.~\cite{BBKT} is
$\zeta_{3}^A = \zeta^\parallel_{3}$, $\zeta_{3}^V =
\omega_{3}^\parallel/14$, 
$\zeta_{3}^T = \omega_{3}^\perp/14$, $\zeta_3^\parallel\omega^A_{1,0} =
\widetilde\omega_{3}^\parallel$; G-parity breaking terms were not
considered in Ref.~\cite{BBKT}. For equal mass quarks,
$\Phi^{\perp,\parallel}_{3;K^*}$ are antisymmetric under
$\alpha_1\leftrightarrow \alpha_2$, whereas 
$\widetilde\Phi^{\parallel}_{3;K^*}$ is symmetric.

All these parameters can be defined in terms of
matrix elements of local twist-3 operators. For chiral-odd operators,
for instance, one has
\begin{eqnarray}
\langle 0 | \bar q \sigma_{z\xi} gG_{z\xi}
s|K^*(P,\lambda)\rangle & = & f_{K^*}^\perp m_{K^*}^2 
(e^{(\lambda)}z)(pz)\kappa_{3K^*}^\perp\,, \nonumber\\
\langle 0 | \bar q \sigma_{z\xi}  [iD_z,gG_{z\xi}] s
- \frac{3}{7}\, i\partial_z \bar q \sigma_{z\xi} 
gG_{z\xi} s | K^*(P,\lambda)\rangle
& = & f_{K^*}^\perp m_{K^*}^2 (e^{(\lambda)}z)(pz)^2
\,\frac{3}{28}\, \lambda_{3K^*}^\perp\,,
\nonumber\\
\langle 0 | \bar q i \!\stackrel{\leftarrow}{D}_z\!
\sigma_{z\xi}  gG_{z\xi} s - \bar q
\sigma_{z\xi}  gG_{z\xi} i\! \stackrel{\rightarrow}{D}_z\! s
  | K^*(P,\lambda)\rangle & = & f_{K^*}^\perp m_{K^*}^2 (e^{(\lambda)}z)(pz)^2
\,\frac{1}{14}\, \omega_{3K^*}^\perp\,;\label{3.12}
\end{eqnarray}
the formulas for chiral-even operators are analogous. 
Numerical values for these parameters can be obtained from QCD sum 
rules and will be discussed in Section~\ref{sec:4}. 

Of the parameters in (\ref{3.11}), $\zeta_{3K^*}^\parallel$,
$\kappa_{3K^*}^{\perp,\parallel}$, $\omega_{3K^*}^\perp$ and
$\lambda_{3K^*}^\perp$ renormalise multiplicatively in
the chiral limit, the others 
mix with each other. For non-zero strange quark mass, there is
additional mixing with twist-2 parameters. Here,
we write down explicitly only the RG-improved relations for the above 
5 parameters; a full discussion, including also $\lambda_3^\parallel$,
$\omega_3^\parallel$, is given in App.~\ref{app:A}.
The relations can be written in compact form as
\begin{equation}
P_i(\mu^2) = L^{(\gamma_P)_i/\beta_0}\, P_i(\mu_0^2) + \sum_{j=1}^3
C_{ij} \left( L^{(\gamma_Q)_{ij}/\beta_0} -
L^{(\gamma_P)_i/\beta_0}\right) Q_{ij}(\mu_0^2)\label{xx}
\end{equation}
with the LO scaling factor $L=\alpha_s(\mu^2)/\alpha_s(\mu_0^2)$. If
there is a flavour threshold $\mu_{\rm th}$ 
between $\mu_0$ and $\mu$ changing the number of active flavours from
$n_f$ to $n_f+1$, then one has to replace 
$$L^{1/\beta_0}\to
(\alpha(\mu^2)/\alpha(\mu^2_{\rm th}))^{1/\beta_0(n_f+1)}(\alpha(\mu^2_{\rm
  th})/\alpha(\mu_0^2))^{1/\beta_0(n_f)}.
$$ 
The parameters in (\ref{xx}) are given by:
\begin{eqnarray}
P & = & \{f_{K^*}^\parallel \zeta_{3K^*}^\parallel,\,f_{K^*}^\parallel
\kappa_{3K^*}^\parallel,\,f_{K^*}^\perp \kappa_{3K^*}^\perp,\,f_{K^*}^\perp
\omega_{3K^*}^\perp,\,f_{K^*}^\perp \lambda_{3K^*}^\perp\}\,,\nonumber\\
Q_{1(2)} &=& \frac{f_{K^*}^\perp}{m_{K^*}}\,\{
m_s\pm m_q,\,(m_s\mp m_q) a_1^\perp,\,(m_s\pm m_q)a_2^\perp\}\,,\nonumber\\
Q_{3,5} &=& \frac{f_{K^*}^\parallel}{m_{K^*}}\,\{
m_s-m_q,\,(m_s+m_q) a_1^\parallel,\,(m_s-m_q) a_2^\parallel\}\,,\nonumber\\
Q_{4}& =& \frac{f_{K^*}^\parallel}{m_{K^*}}\,\{
m_s+m_q,\,(m_s-m_q) a_1^\parallel,\,(m_s+m_q) a_2^\parallel\}\,,
\nonumber\\
\gamma_P & = & \left\{
\frac{77}{9},\,\frac{77}{9},\,\frac{55}{9},\,\frac{73}{9},\,
\frac{104}{9}\right\}, \nonumber\\
(\gamma_Q)_{1,2} & = & \left\{ \frac{16}{3},\, 8,\,
\frac{88}{9} \right\},\qquad
(\gamma_Q)_{3,4,5}  =  \left\{ 4,\, \frac{68}{9},\,
\frac{86}{9} \right\},\nonumber\\
C & = & \left(
\begin{array}{r@{\hskip10pt}r@{\hskip10pt}r}
\ds \frac{2}{29} & \ds\frac{6}{25} & 0\\[10pt]
\ds -\frac{2}{29} & \ds-\frac{6}{25} & 0\\[10pt]
\ds-\frac{4}{19} & \ds \frac{12}{65} & 0\\[10pt]
\ds \frac{14}{37} & \ds-\frac{42}{25} & \ds\frac{12}{13}\\[10pt]
\ds -\frac{1}{85} & \ds -\frac{1}{5} & \ds-\frac{4}{15}
\end{array}
\right).
\end{eqnarray}
Implicit formulas for the remaining 4 parameters in (\ref{3.11}) can
be found in App.~\ref{app:A}. Numerical values will be given in the
next section.

Using (\ref{3.11}), and the corresponding relations
for twist-2 DAs, one obtains expressions for the twist-3
two-particle DAs, which are valid to NLO
in the conformal expansion. As discussed in Ref.~\cite{BBKT}, the
structure of this expansion is complicated by the fact that 
these DAs do not correspond to a fixed projection of the quark fields' 
Lorentz-spin $s$. The resulting expansion is in $C^{3/2}_n(\xi)$ for
$\psi^{\perp,\parallel}_{3;K^*}$ and $C^{1/2}_n(\xi)$ for
$\phi^{\perp,\parallel}_{3;K^*}$:
\begin{eqnarray}
\phi_{3;K^*}^\parallel(u) & = & 3\xi^2 + \frac{3}{2}\,\xi(3\xi^2-1)
a_1^\perp + \frac{3}{2}\,\xi^2 (5\xi^2-3)a_2^\perp\nonumber\\
&&{} + \left(\frac{15}{2}\,\kappa_{3K^*}^\perp -
\frac{3}{4}\,\lambda_{3K^*}^\perp\right) \xi (5\xi^2-3) +
\frac{5}{8}\,\omega_{3K^*}^\perp (3-30\xi^2 + 35\xi^4) \nonumber\\
&&{}+\frac{3}{2}\,\frac{m_s+m_q}{m_{K^*}}\,
\frac{f^\parallel_{K^*}}{f_{K^*}^\perp}
\left\{ 1 + 8\xi a_1^\parallel + 3 (7-30 u\ub )a_2^\parallel + \xi \ln
\ub (1+3 a_1^\parallel + 6 a_2^\parallel)\right.\nonumber\\
&&{}\left.\hspace*{3.6cm} - \xi \ln u (1-3
a_1^\parallel + 6 a_2^\parallel)\right\}\nonumber\\
&&{}-\frac{3}{2}\,\frac{m_s-m_q}{m_{K^*}}\,
\frac{f^\parallel_{K^*}}{f_{K^*}^\perp}\,
\xi \left\{ 2 + 9\xi a_1^\parallel + 2 (11-30 u\ub )a_2^\parallel + \ln
\ub (1+3 a_1^\parallel + 6 a_2^\parallel)\right.\nonumber\\
&&{}\hspace*{3.9cm}\left. + \ln u (1-3
a_1^\parallel + 6 a_2^\parallel)\right\},\\
\psi_{3;K^*}^\parallel(u) & = & 6 u\bar u \left\{ 1 + \left(
\frac{a_1^\perp}{3} + \frac{5}{3} \kappa_{3K^*}^\perp\!\right)
C_1^{3/2}(\xi) + \left( \frac{a_2^\perp}{6}  + \frac{5}{18}
\omega_{3K^*}^\perp\right) C_2^{3/2}(\xi) -
\frac{1}{20}\lambda_{3K^*}^\perp C_3^{3/2}(\xi)\right\}\nonumber\\
&&{}+ 3\,
\frac{m_s+m_q}{m_{K^*}}\,\frac{f_{K^*}^\parallel}{f_{K^*}^\perp}\left\{
u\bar u (1 + 2 \xi a_1^\parallel + 3 (7-5u\bar u) a_2^\parallel) + \bar
u \ln \bar u (1+3 a_1^\parallel + 6 a_2^\parallel)\right.\nonumber\\
&&{}\left.\hspace*{3.2cm} + u \ln u 
(1-3 a_1^\parallel + 6 a_2^\parallel)\right\}\nonumber\\
&&- 3\,
\frac{m_s-m_q}{m_{K^*}}\,\frac{f_{K^*}^\parallel}{f_{K^*}^\perp}\left\{
u\bar u (9 a_1^\parallel + 10\xi a_2^\parallel) + \bar
u \ln \bar u (1+3 a_1^\parallel + 6 a_2^\parallel)\right.\nonumber\\
&&{}\left.\hspace*{3.2cm} - u \ln u 
(1-3 a_1^\parallel + 6 a_2^\parallel)\right\},\\
\psi_{3;K^*}^\perp(u) & = & 6 u\bar u \left\{ 1 + \left(
\frac{1}{3} a_1^\parallel + \frac{20}{9} \kappa_{3K^*}^\parallel\right)
C_1^{3/2}(\xi) 
\right.\nonumber\\
&&\left.\hspace*{0pt} + \left( \frac{1}{6} a_2^\parallel +
\frac{10}{9}\zeta_{3K^*}^\parallel + \frac{5}{12}\,
\omega_{3K^*}^\parallel-\frac{5}{24}\,
\widetilde\omega_{3K^*}^\parallel\!\right)
C_2^{3/2}(\xi) +
\left(\frac{1}{4}\widetilde\lambda_{3K^*}^\parallel - \frac{1}{8}
\lambda_{3K^*}^\parallel\right) C_3^{3/2}(\xi)\right\}\nonumber\\
&&{}+ 6\,
\frac{m_s+m_q}{m_{K^*}}\,\frac{f_{K^*}^\perp}{f_{K^*}^\parallel}\left\{
u\bar u (2 + 3 \xi a_1^\perp + 2 (11-10u\bar u) a_2^\perp) + \bar
u \ln \bar u (1+3 a_1^\perp + 6 a_2^\perp)\right.\nonumber\\
&&{}\left.\hspace*{3.2cm} + u \ln u 
(1-3 a_1^\perp + 6 a_2^\perp)\right\}\nonumber\\
&&- 6\,
\frac{m_s-m_q}{m_{K^*}}\,\frac{f_{K^*}^\perp}{f_{K^*}^\parallel}\left\{
u\bar u (9 a_1^\perp + 10\xi a_2^\perp) + \bar
u \ln \bar u (1+3 a_1^\perp + 6 a_2^\perp)\right.\nonumber\\
&&{}\left.\hspace*{3.2cm} - u \ln u 
(1-3 a_1^\perp + 6 a_2^\perp)\right\},\\
\phi_{3;K^*}^\perp(u) & = & \frac{3}{4}\,(1+\xi^2) + \frac{3}{2}\,\xi^3
a_1^\parallel + \left\{ \frac{3}{7}\,a_2^\parallel + 5
\zeta_{3K^*}^\parallel \right\} (3\xi^2-1) + \left\{5\kappa_{3K^*}^\parallel
- \frac{15}{16}\,\lambda_{3K^*}^\parallel \right.\nonumber\\
&&{} \left. +
\frac{15}{8}\,\widetilde{\lambda}_{3K^*}^\parallel \right\} \xi
(5\xi^2-3)+ \left\{ \frac{9}{112}\,a_2^\parallel +
\frac{15}{32}\,\omega_{3K^*}^\parallel -
\frac{15}{64}\,\widetilde\omega_{3K^*}^\parallel \right\} (35\xi^4-30
\xi^2+3)\nonumber\\
&&{}+\frac{3}{2}\,\frac{m_s+m_q}{m_{K^*}}\,\frac{f_{K^*}^\perp}{
f_{K^*}^\parallel} \left\{ 2 + 9 \xi a_1^\perp + 2 (11-30 u \ub)
    a_2^\perp\right.\nonumber\\
&&{}\left.\hspace*{3.5cm} + 
(1-3 a_1^\perp + 6 a_2^\perp) \ln u + (1+3 a_1^\perp + 6 a_2^\perp) 
\ln \ub \right\}\nonumber\\
&&
    {}-\frac{3}{2}\,\frac{m_s-m_q}{m_{K^*}}\,\frac{f_{K^*}^\perp}{
f_{K^*}^\parallel} \left\{ 2\xi + 9 (1-2u \ub) a_1^\perp + 2 \xi (11-20 u \ub)
    a_2^\perp\right.\nonumber\\
&&{}\left.\hspace*{3.5cm} + 
(1+3 a_1^\perp + 6 a_2^\perp) \ln \ub - (1-3 a_1^\perp + 6 a_2^\perp) 
\ln u \right\}.
\end{eqnarray}
These expressions are our final results for the two-particle twist-3
DAs and supersede those given in Ref.~\cite{BBKT} where G-parity
violating terms in $\kappa_3$ and $\lambda_3$ were not included. The terms
multiplying $\ln u$ and $\ln\bar u$ are the first three terms in the
conformal expansion of $d/(du) \phi^{\parallel,\perp}_{2;K^*}(0)$ and
$d/(du)  \phi^{\parallel,\perp}_{2;K^*}(1)$, respectively.
Numerical values for the hadronic parameters are given in Tab.~\ref{tab:num}.

\section{Models for Distribution Amplitudes}\label{sec:4}
\setcounter{equation}{0}

\begin{table}[tb]
\renewcommand{\arraystretch}{1.3}
\addtolength{\arraycolsep}{3pt}
$$
\begin{array}{l || c | c ||  l | l || c | c }
\hline
& \multicolumn{2}{c||}{\rho} & \multicolumn{2}{c||}{K^*}  &  
\multicolumn{2}{c}{\phi}\\
\cline{2-7}
& \mu = 1\,{\rm GeV} & \mu = 2\,{\rm GeV} & \mu = 1\,{\rm GeV} & \mu =
2\,{\rm GeV} & \mu = 1\,{\rm GeV} & \mu = 2\,{\rm GeV}\\
\hline
a_1^\parallel & 0 & 0 & \phantom{-}0.03(2) & \phantom{-}0.02(2) & 0 & 0 
\\
a_1^\perp & 0 & 0 & \phantom{-}0.04(3) & \phantom{-}0.03(3) & 0 & 0
\\
a_2^\parallel & 0.15(7) & 0.10(5) & \phantom{-}0.11(9) & 
\phantom{-}0.08(6) & 0.18(8) & 0.13(6)
\\
a_2^\perp & 0.14(6) & 0.11(5) & \phantom{-}0.10(8) &
\phantom{-}0.08(6) & 0.14(7) & 0.11(5)
\\\hline
\zeta_{3V}^\parallel 
& 0.030(10) & 0.020(9) & \phantom{-}0.023(8) & \phantom{-}0.015(6) & 
0.024(8) & 0.017(6)
\\
\widetilde\lambda_{3V}^\parallel 
& 0 & 0 & \phantom{-}0.035(15)& \phantom{-}0.017(8) & 0 & 0
\\
\widetilde\omega_{3V}^\parallel 
& -0.09(3) & -0.04(2) & -0.07(3) &  -0.03(2) & -0.045(15) & -0.022(8)
\\
\kappa_{3V}^\parallel 
& 0 & 0 & \phantom{-}0.000(1) & -0.001(2) & 0 & 0
\\
\omega_{3V}^\parallel 
& 0.15(5) & 0.09(3) & \phantom{-}0.10(4) & \phantom{-}0.06(3) &
0.09(3) & 0.06(2)
\\
\lambda_{3V}^\parallel 
& 0 & 0 & -0.008(4) & -0.004(2) & 0 & 0
\\
\kappa_{3V}^\perp 
& 0 & 0 & \phantom{-}0.003(3) & -0.001(2) & 0 & 0 
\\
\omega_{3V}^\perp 
& 0.55(25) & 0.37(19) & \phantom{-}0.3(1) & \phantom{-}0.2(1) &  
0.20(8) & 0.15(7)
\\
\lambda_{3V}^\perp 
& 0 & 0 & -0.025(20) & -0.015(10) & 0 & 0\\\hline 
\end{array}
$$
\renewcommand{\arraystretch}{1}
\addtolength{\arraycolsep}{-3pt}
\vspace*{-10pt}
\caption[]{\small Twist-2 and -3 hadronic parameters 
at the scale  $\mu= 1\,{\rm GeV}$ and scaled to up $\mu= 2\,{\rm GeV}$, 
using the evolution equations
(\ref{xx}). The sign of the twist-3 parameters
corresponds to the sign convention for the strong coupling defined by
the covariant derivative $D_\mu = \partial_\mu - i g A^a_\mu t^a$; they
change sign if $g$ is fixed by $D_\mu = \partial_\mu + i g
A^a_\mu t^a$.}\label{tab:num}
\end{table}

In this section we compile the numerical estimates of all necessary parameters 
and present explicit models of the twist-3 two-particle
distribution amplitudes introduced in the last section.
The important point is that these DAs are related to three-particle
ones by exact
QCD equations of motion and have to be used together; this
guarantees the consistency of the approximation.
Our approximation thus introduces a minimum number of non-perturbative 
parameters, which are defined as  matrix elements of certain local operators 
between the vacuum and the meson state, and which we estimate using
QCD sum rules. More
sophisticated models can be constructed in a systematic way by adding 
contributions of higher conformal partial waves when estimates of the relevant 
non-perturbative matrix elements will become available. 

Our  approach involves the implicit assumption that the conformal partial 
wave expansion is well convergent. This can be justified rigorously 
at large scales, since the anomalous dimensions of all involved 
operators increase logarithmically with the conformal spin $J$, but 
is non-trivial at relatively low scales of order $\mu \sim (1$--$2)\,$GeV
which we choose as reference scale. 
%

Since orthogonal polynomials of high orders are rapidly oscillating 
functions, a truncated expansion in conformal partial waves is, almost 
necessarily,
oscillatory as well. Such a behaviour is clearly unphysical,
but this does not constitute a real problem since  physical observables 
are given by convolution integrals of distribution amplitudes with
smooth coefficient functions. 
A classical example for this
feature is the $\gamma\gamma^*$-meson form factor, which is governed
by the quantity
$$
\int du\,\frac{1}{u}\, \phi(u) \sim \sum a_i,
$$
where the coefficients $a_i$ are exactly the ``reduced matrix elements''
in the conformal expansion.
The oscillating terms are averaged over and strongly suppressed. 
Stated otherwise: 
models of distribution amplitudes should generally be understood as
distributions (in the mathematical sense).    

We give all relevant numerical
input parameters for our model DAs in
Table~\ref{tab:num}, at the scale $\mu=1\,$GeV, which is appropriate for
QCD sum-rule results, and, using the LO and NLO scaling relations
given in Secs.~\ref{sec:2.3} and \ref{sec:3}, at the scale $\mu=2\,$GeV,
in order to facilitate the comparison with future lattice
determinations of these quantities. The mixing of $K^*$ and $\phi$ parameters
with operators of lower twist depending on $m_s$ is numerically
small. In evaluating the sum rules, we have chosen the values of the
continuum threshold $s_0$ as given in App.~\ref{app:C}. The sum rules
are actually rather insensitive to that parameter, due to the
smallness of the perturbative contribution, but are not very stable in
the Borel parameter $M^2$, which is the reason for the large
uncertainties in Tab.~\ref{tab:num}. The biggest contribution to
G-conserving parameters comes from the gluon condensate. For
G-breaking parameters, on the other hand, this contribution is
suppressed by a factor $m_s^2$ and, as a consequence, all G-breaking
parameters are considerably smaller than the G-conserving ones. The
table also shows that SU(3) breaking is relevant for all parameters.

At this point we would like to compare our results with those
available in the literature. We have discussed the twist-2 parameters
already in Sec.~{2.3}. As for twist-3, only results for $\rho$ are
available. A long time ago, the chiral-even parameters were
determined, in Ref.~\cite{CZ}, as
\begin{equation}
\zeta^\parallel_{3\rho}(1\,{\rm GeV}) = 0.033\pm0.003\,,\quad
\omega^\parallel_{3\rho}(1\,{\rm GeV}) = 0.2\,,\quad
\tilde\omega^\parallel_{3\rho}(1\,{\rm GeV}) = -0.1\,.
\end{equation}
A comparison with Tab.~\ref{tab:num} shows that these values agree
quite well with ours, although we think that the uncertainty of
$\zeta^\parallel_{3\rho}$ was underestimated in \cite{CZ}. A value for
$\omega^\perp_{3\rho}$ was obtained in Ref.~\cite{BBKT}:
$\omega^\perp_{3\rho}(1\,{\rm GeV}) = 0.3\pm 0.3$, which is a bit
smaller than our result. 

\begin{figure}[tb]
$$
\epsfxsize=0.48\textwidth\epsffile{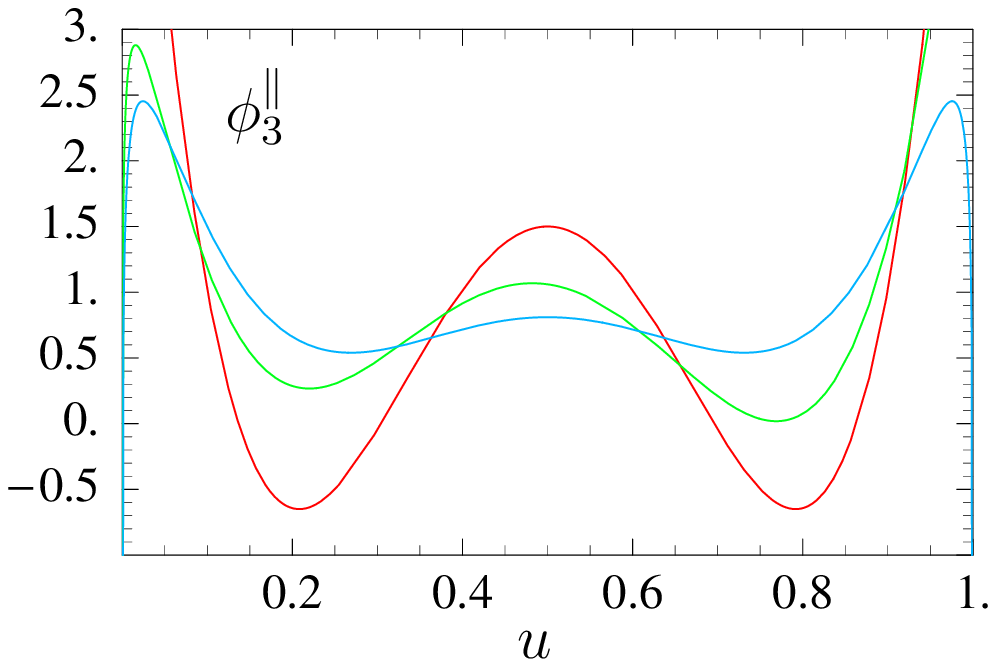}\quad
\epsfxsize=0.48\textwidth\epsffile{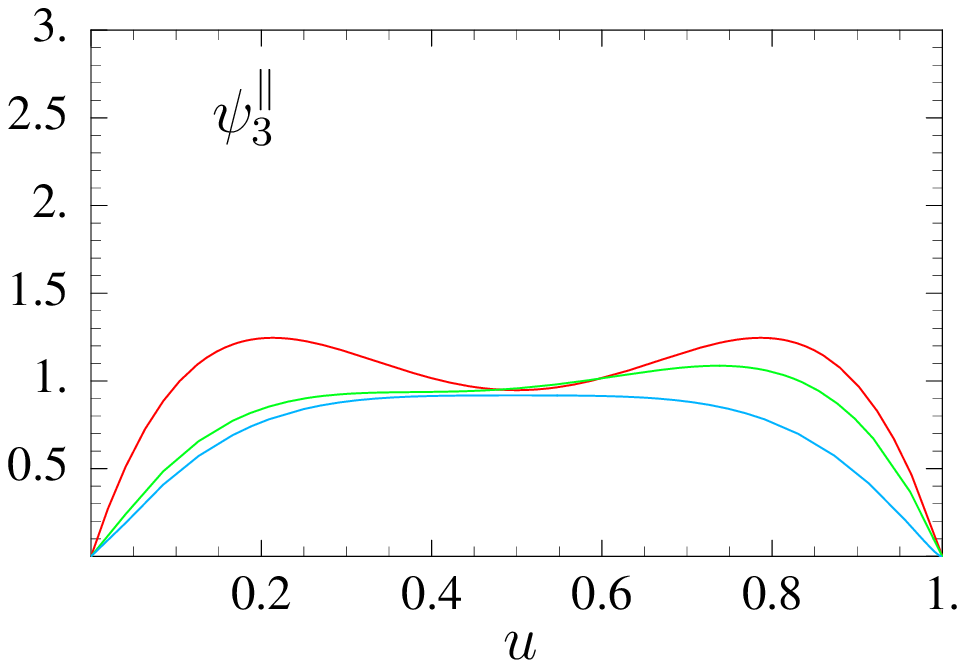}
$$
\vskip-15pt
\caption[]{\small Left panel: $\phi^\parallel_{3}$ as a function of
  $u$ for the central values of hadronic parameters, for $\mu=1\,$GeV.
  Red line: $\phi_{3;\rho}^\parallel$, green:
  $\phi_{3;K^*}^\parallel$, blue: $\phi_{3;\phi}^\parallel$. Right
  panel: same for $\psi^\parallel_{3}$.}\label{fig:1}
$$
\epsfxsize=0.48\textwidth\epsffile{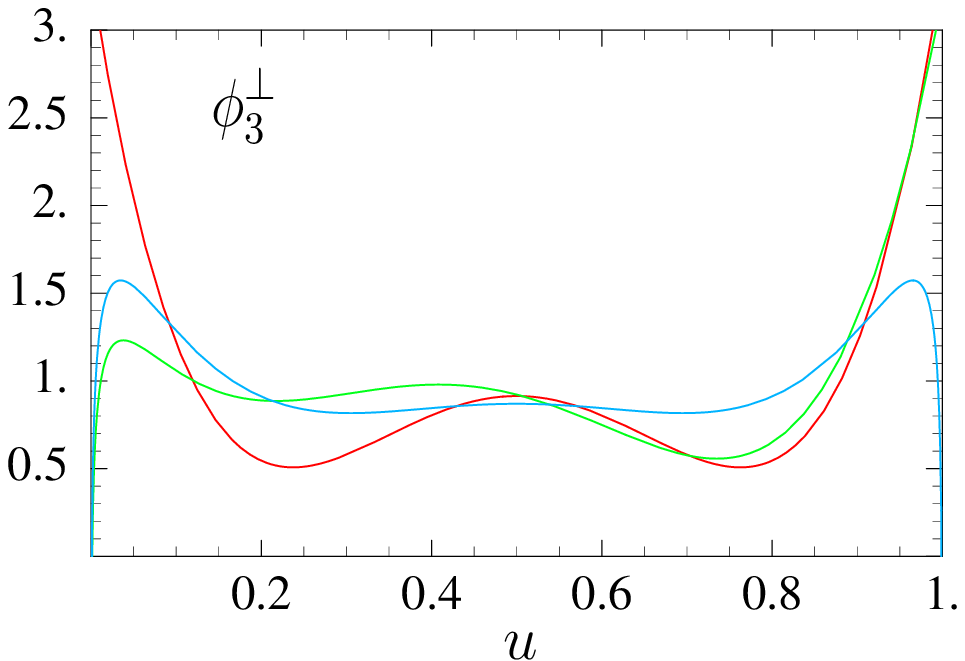}\quad
\epsfxsize=0.48\textwidth\epsffile{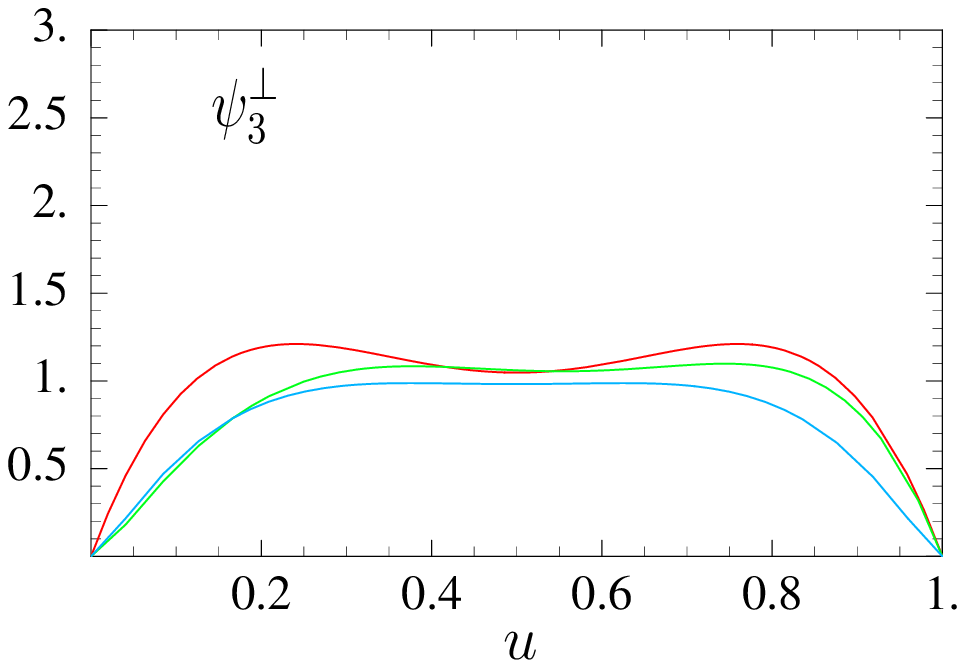}
$$
\vskip-15pt
\caption[]{\small Left panel: $\phi^\perp_{3}$ as a function of
  $u$ for the central values of hadronic parameters, for $\mu=1\,$GeV.
  Red line: $\phi_{3;\rho}^\perp$, green:
  $\phi_{3;K^*}^\perp$, blue: $\phi_{3;\phi}^\perp$. Right
  panel: same for $\psi^\perp_{3}$.}\label{fig:2}
\end{figure}
In Fig.~\ref{fig:1} we plot the
longitudinal twist-3 two-particle DAs $\phi^\parallel_{3}$ and 
$\psi^\parallel_{3}$ for the $\rho$ meson, assuming massless quarks,
and for the $K^*$ and $\phi$ mesons, 
together with the corresponding asymptotic DAs. Figure~\ref{fig:2}
shows the transverse DAs $\phi^\perp_{3}$ and $\psi^\perp_{3}$.
The figures show that quark-mass corrections significantly modify
 the end-point behaviour of $\phi_3^{\parallel,\perp}$, where they induce
a logarithmic end-point divergency, even if the contributions of
gluonic operators are
 neglected. This is not a problem because, as mentioned above, the DAs
 themselves need not be finite, it is only their convolution with
 perturbative scattering amplitudes that is meaningful. The figures
 also show that the effect of SU(3) breaking (the difference between
 the red and the other curves) is quite pronounced for
 all DAs, whereas the G-parity breaking terms (the asymmetry of the
 green curve) have only minor impact, which is due to the numerical 
smallness of the corresponding hadronic parameters.

\section{Summary and Conclusions}\label{sec:5}
\setcounter{equation}{0}

In this paper we have studied the twist-3 two- and three-particle
distribution amplitudes of $K^*$ and $\phi$ mesons in QCD and 
expressed them in a
model-independent way by a minimal number of non-perturbative
parameters. The work presented here is an extension of
Refs.~\cite{BBL06,BBKT} 
and completes the analysis of  SU(3)-breaking
in vector meson distribution amplitudes to twist-3 accuracy. 
Our approach consists of two components.
One is the use of the QCD
equations of motion, which allow dynamically dependent
DAs to be expressed in terms of independent ones. The other ingredient
is conformal
expansion, which makes it possible to separate transverse and
longitudinal variables in the wave functions, the former ones being
governed by renormalisation-group equations, the latter ones being
described in terms of irreducible representations of the corresponding
symmetry group.
We have derived expressions for all twist-3 two- and
three-particle distribution amplitudes to next-to-leading order in the
conformal expansion, including both chiral corrections ${\mathcal
  O}(m_s+m_q)$ and G-parity-breaking corrections ${\mathcal
  O}(m_s-m_q)$; the corresponding formulas are given in
Sec.~\ref{sec:3}.

We have also done a complete reanalysis of the numerical values of the
relevant twist-3 hadronic parameters from QCD sum rules.
Our sum rules can be compared, in the chiral limit, with
existing calculations for the $\rho$ \cite{BBKT,CZ}. 
We have also studied the scale-dependence of all
parameters to leading-logarithmic, or, if possible,
next-to-leading-logarithmic accuracy, taking into account the 
mixing with operators
depending on the strange-quark mass $m_s$. 
Our final numerical results, at the scales 1 and 2~GeV, are collected
in Tab.\ref{tab:num}. 

Preliminary versions of our results have already been applied in
studies of $B\to (\rho,K^*)\gamma$ decays
\cite{BVgamma,BJZ,frueher}. These processes are also sensitive to
twist-4 distribution amplitudes which we will study in a separate
publication. While the parametrisations given in Sec.~\ref{sec:3} are
general, the actual numerical values for hadronic parameters given in
Sec.~\ref{sec:2.3} and Tab.~\ref{tab:num} are obtained within the
framework of QCD sum rules. We are looking forward to a confirmation
of these values from independent non-perturbative methods, for
instance lattice QCD.

\section*{Acknowledgements}

G.W.~Jones acknowledges a PPARC student fellowship. This work was
supported in part by the EU network
contract No.\ MRTN-CT-2006-035482, {\sc Flavianet}.

\appendix

\section*{Appendices}
\renewcommand{\theequation}{\Alph{section}.\arabic{equation}}
\renewcommand{\thetable}{\Alph{table}}
\setcounter{section}{0}
\setcounter{table}{0}

\section{Scale-Dependence of Twist-3 Parameters}\label{app:A}

In this appendix we derive the renormalisation-group improved
expression (\ref{xx}) for twist-3 parameters and also give an implicit
relation for the scaling of the chiral-even parameters $\omega_3$, 
$\tilde\omega_3$, $\lambda_3$ and $\tilde\lambda_3$.

Let us introduce the following notation for the relevant
quark-quark-gluon operators, see Eq.~(\ref{3.11}):
\begin{eqnarray}\label{yy}
 O_3^T(z,vz,-z) &=&
\bar q(z)\sigma_{z\nu} g G_{z\nu}(vz) s(-z)\,,\nonumber\\
O_3(z,vz,-z) &=& \bar q(z) g G_{\perp z}(vz)i  \gamma_z s(-z),\qquad 
\tilde O_3(z,vz,-z) =  \bar q(z) g \tilde 
G_{\perp z}(vz)\gamma_z\gamma_5 s(-z)\,.\hspace*{25pt}
\end{eqnarray}
In principle it is possible to establish evolution equations for these
non-local operators using the light-ray operator technique developed in
Ref.~\cite{string}. The scale-dependence of the parameters in
(\ref{3.11}) then follows from a projection of the evolution equation
on the corresponding conformal wave. Another approach is to make use
of the results derived in the literature 
for the anomalous dimensions of moments of the corresponding nucleon
structure functions. The explicit relations between these anomalous
dimensions, given in Ref.~\cite{andim}, 
and those needed for our twist-3 parameters
have been established in Ref.~\cite{BBKT}. 
Neglecting quark mass corrections, one has, in the notations of
Ref.~\cite{BBKT}, 
\begin{eqnarray}
[f^\perp_{K^*}\kappa_{3K^*}^\perp](\mu^2) = L^{\Gamma^{T+}_2/\beta_0} 
[f^\perp_{K^*}\kappa_{3K^*}^\perp](\mu_0^2)\,,&&
[f^\perp_{K^*}\omega_{3K^*}^\perp](\mu^2) =  L^{\Gamma^{T+}_3/\beta_0} 
[f^\perp_{K^*}\omega_{3K^*}^\perp](\mu_0^2)\,,\nonumber\\
{}[f^\perp_{K^*}\lambda_{3K^*}^\perp](\mu^2) = L^{\Gamma^{T-}_3/\beta_0} 
[f^\perp_{K^*}\lambda_{3K^*}^\perp](\mu_0^2)\,,\nonumber\\
f^\parallel_{K^*}\zeta^\parallel_{3K^*}(\mu^2) =  L^{\Gamma^-_2/\beta_0}
f^\parallel_{K^*}\zeta^\parallel_{3K^*}(\mu_0^2)\,,&&
f^\parallel_{K^*}\kappa^\parallel_{3K^*}(\mu^2) =  L^{\Gamma^+_2/\beta_0}
f^\parallel_{K^*}\kappa^\parallel_{3K^*}(\mu_0^2)\,,\nonumber\\
f^\parallel_{K^*}\left(\begin{array}{l}
\ds\frac{2}{3}\,\omega^\parallel_{3K^*} - \tilde\omega^\parallel_{3K^*}\\[10pt]
\ds\frac{2}{3}\,\omega^\parallel_{3K^*} + \tilde\omega^\parallel_{3K^*}
\end{array}\right)^{\mu^2} 
& = & L^{\Gamma_3^+/\beta_0} f^\parallel_{K^*}
\left(\begin{array}{l}
\ds\frac{2}{3}\,\omega^\parallel_{3K^*} - \tilde\omega^\parallel_{3K^*}\\[10pt]
\ds\frac{2}{3}\,\omega^\parallel_{3K^*} + \tilde\omega^\parallel_{3K^*}
\end{array}\right)^{\mu_0^2}
\,,\nonumber\\
f^\parallel_{K^*}\left(\begin{array}{l}
\ds\frac{2}{3}\,\tilde\lambda^\parallel_{3K^*} + \lambda^\parallel_{3K^*}\\[10pt]
\ds\frac{2}{3}\,\tilde\lambda^\parallel_{3K^*} - \lambda^\parallel_{3K^*}
\end{array}\right)^{\mu^2} 
&=&  L^{\Gamma_3^-/\beta_0} f^\parallel_{K^*}
\left(\begin{array}{l}
\ds\frac{2}{3}\,\tilde\lambda^\parallel_{3K^*} + \lambda^\parallel_{3K^*}\\[10pt]
\ds\frac{2}{3}\,\tilde\lambda^\parallel_{3K^*} - \lambda^\parallel_{3K^*}
\end{array}\right)^{\mu_0^2}\,,\label{A.2}
\end{eqnarray}
where $L$ is the leading-log scaling factor
$L=\alpha_s(\mu^2)/\alpha_s(\mu_0^2)$. The factor $2/3$ comes from the
relative factors in (\ref{3.12}). The anomalous dimensions are
given by 
\begin{eqnarray}
\Gamma^{T+}_2 &  = & C_A + \frac{7}{3}\,C_F\,,\quad
\Gamma^{T+}_3 = C_A + \frac{23}{6}\, C_F\,,\quad
\Gamma^{T-}_3 = \frac{10}{3}\,C_A + \frac{7}{6}\,C_F\,,\nonumber\\
\Gamma^+_2 & = & 3\, C_A - \frac{1}{3}\,C_F = \Gamma_2^-\,,\nonumber\\
\Gamma_3^+ & = & \left(\begin{array}{l@{\quad}l}
\ds\phantom{-}\frac{7}{3}\, C_A + \frac{8}{3}\,C_F & 
\ds -\frac{2}{3}\,C_A + \frac{2}{3}\,C_F\\[10pt]
\ds -\frac{4}{3}\,C_A + \frac{5}{3}\,C_F & 
\ds\phantom{-}4C_A +\frac{1}{6}\,C_F
\end{array}\right),\nonumber\\
\Gamma_3^- & = & \left(\begin{array}{l@{\quad}l}
\ds\phantom{-}4\, C_A + \frac{1}{6}\,C_F & 
\ds -\frac{4}{3}\,C_A + \frac{5}{3}\,C_F\\[10pt]
\ds -\frac{2}{3}\,C_A + \frac{2}{3}\,C_F & 
\ds\phantom{-}\frac{7}{3}\,C_A +\frac{8}{3}\,C_F
\end{array}\right).
\end{eqnarray}

For massive quarks, the scaling relations receive corrections in
$m_s\pm m_q$, depending on the G-parity of the parameter. These
corrections are induced by mixing of the operators in (\ref{yy}) with
twist-2 operators and can be calculated using the
light-ray-operator technique of Ref.~\cite{string} resulting in the
following compact expressions:
\begin{eqnarray}
   O_3^T(z,vz,0)^{\mu^2} &=& O_3^T(z,vz,0)^{\mu_0^2} + 
i\,\frac{C_F\alpha_s}{2\pi}\,\ln\,\frac{\mu_0^2}{\mu^2}
\int_0^1 dt \left\{\frac{m_s}{v}\, \Big[O_2(z,vz) - 2 
t O_2(z,tvz)\Big]\right.\nonumber\\
&&\left. + \frac{m_q}{1-v}\, \Big[O_2(vz,0) - 2 
t O_2((1-(1-v)t)z,0)\Big]\right\} + \dots,\nonumber\\
O_3(z,vz,0)^{\mu^2} &=&   O_3(z,vz,0)^{\mu_0^2} + 
i\,\frac{C_F\alpha_s}{4\pi}\,\ln\,\frac{\mu_0^2}{\mu^2}
\int_0^1 dt \left\{\frac{m_s}{v}\, \Big[O_2^T(z,vz) - 2 
t O_2^T(z,tvz)\Big]\right.\nonumber\\
&&\left. + \frac{m_q}{1-v}\, \Big[O_2^T(vz,0) - 2 
t O_2^T((1-(1-v)t)z,0)\Big]\right\} + \dots,\nonumber\\
\tilde O_3(z,vz,0)^{\mu^2} &=&   \tilde O_3(z,vz,0)^{\mu_0^2}
 + i\,\frac{C_F\alpha_s}{4\pi}\,\ln\,\frac{\mu_0^2}{\mu^2}
\int_0^1 dt \left\{-\frac{m_s}{v}\, \Big[O_2^T(z,vz) - 2 
t O_2^T(z,tvz)\Big]\right.\nonumber\\
&&\left. + \frac{m_q}{1-v}\, \Big[O_2^T(vz,0) - 2 
t O_2^T((1-(1-v)t)z,0)\Big]\right\} + \dots\label{A.4}
\end{eqnarray}
The twist-2 operators in the above relations are given by
\begin{equation}
    O_2(az,bz) = \bar q(az)\gamma_z s(bz)\,,\qquad
O_2^T(az,bz) = \bar q(az)\sigma_{\perp z} s(bz)
\end{equation}
and the dots stand for $O(\alpha_s)$ contributions from the twist-3
operators.

Taking (\ref{A.2}) and (\ref{A.4}) together, one finds that the G-even
parameters mix with $$f_{K^*}^{\perp(\parallel)} \{ (m_s+m_q),
(m_s-m_q) a_1^{\perp(\parallel)}, (m_s+m_q) a_2^{\perp(\parallel)}\},$$
whereas the G-odd ones mix with $$f_{K^*}^{\perp(\parallel)} \{ (m_s-m_q),
(m_s+m_q) a_1^{\perp(\parallel)}, (m_s-m_q)
a_2^{\perp(\parallel)}\}.$$ For the $\perp$ parameters, and 
$\zeta^\parallel_3$ and $\kappa^\parallel_3$, the combination of
(\ref{A.2}) and the projection of (\ref{A.4}) onto the corresponding
partial waves results in the scaling
relations given in (\ref{xx}). For the remaining parameters, one finds
the following relations:
\begin{eqnarray}
\left(
\begin{array}{c}
f_{K^*}^\parallel\omega^\parallel_{3K^*} \\[10pt]
f_{K^*}^\parallel\tilde\omega^\parallel_{3K^*} \\[10pt]
\ds\frac{m_s+m_q}{m_{K^*}}\, f_{K^*}^\perp \\[10pt]
\ds\frac{m_s-m_q}{m_{K^*}}\, f_{K^*}^\perp a_1^\perp(K^*) \\[10pt]
\ds\frac{m_s+m_q}{m_{K^*}}\, f_{K^*}^\perp a_2^\perp(K^*)
\end{array}
\right)^{\mu^2}
= L^{\Gamma_\omega/\beta_0}
\left(
\begin{array}{c}
f_{K^*}^\parallel\omega^\parallel_{3K^*} \\[10pt]
f_{K^*}^\parallel\tilde\omega^\parallel_{3K^*} \\[10pt]
\ds\frac{m_s+m_q}{m_{K^*}}\, f_{K^*}^\perp \\[10pt]
\ds\frac{m_s-m_q}{m_{K^*}}\, f_{K^*}^\perp a_1^\perp(K^*) \\[10pt]
\ds\frac{m_s+m_q}{m_{K^*}}\, f_{K^*}^\perp a_2^\perp(K^*)
\end{array}
\right)^{\mu_0^2}\,,\nonumber\\
\left(
\begin{array}{c}
f_{K^*}^\parallel\lambda^\parallel_{3K^*} \\[10pt]
f_{K^*}^\parallel\tilde\lambda^\parallel_{3K^*} \\[10pt]
\ds\frac{m_s-m_q}{m_{K^*}}\, f_{K^*}^\perp \\[10pt]
\ds\frac{m_s+m_q}{m_{K^*}}\, f_{K^*}^\perp a_1^\perp(K^*) \\[10pt]
\ds\frac{m_s-m_q}{m_{K^*}}\, f_{K^*}^\perp a_2^\perp(K^*)
\end{array}
\right)^{\mu^2}
= L^{\Gamma_\lambda/\beta_0}
\left(
\begin{array}{c}
f_{K^*}^\parallel\lambda^\parallel_{3K^*} \\[10pt]
f_{K^*}^\parallel\tilde\lambda^\parallel_{3K^*} \\[10pt]
\ds\frac{m_s-m_q}{m_{K^*}}\, f_{K^*}^\perp \\[10pt]
\ds\frac{m_s+m_q}{m_{K^*}}\, f_{K^*}^\perp a_1^\perp(K^*) \\[10pt]
\ds\frac{m_s-m_q}{m_{K^*}}\, f_{K^*}^\perp a_2^\perp(K^*)
\end{array}
\right)^{\mu_0^2}\,.
\end{eqnarray}
The anomalous dimension matrices are given by
\begin{eqnarray}
\Gamma_\omega 
& = &
\left(
\begin{array}{r@{\hskip10pt}r@{\hskip10pt}r@{\hskip10pt}r@{\hskip10pt}r}
\ds \frac{179}{18} & \ds \frac{7}{4} & \ds -\frac{7}{9} & \ds
\frac{7}{15} & \ds \frac{2}{3}\\[10pt]
\ds\frac{1}{3} & \ds\frac{77}{6} & \ds-\frac{2}{45} & \ds-\frac{2}{5}
& \ds -\frac{4}{15}\\[10pt]
0 & 0 & \ds\frac{16}{3} & 0 & 0\\[10pt]
0 & 0 & 0 & 8 & 0\\[10pt]
0 & 0 & 0 & 0 & \ds\frac{88}{9}
\end{array}
\right),
\quad
\Gamma_\lambda
 = 
\left(
\begin{array}{r@{\hskip10pt}r@{\hskip10pt}r@{\hskip10pt}r@{\hskip10pt}r}
\ds \frac{77}{6} & \ds \frac{1}{3} & \ds \frac{2}{45} & \ds
\frac{2}{5} & \ds \frac{4}{15}\\[10pt]
\ds\frac{7}{4} & \ds\frac{179}{18} & \ds\frac{7}{9} & \ds-\frac{7}{15}
& \ds -\frac{2}{3}\\[10pt]
0 & 0 & \ds\frac{16}{3} & 0 & 0\\[10pt]
0 & 0 & 0 & 8 & 0\\[10pt]
0 & 0 & 0 & 0 & \ds\frac{88}{9}
\end{array}
\right).
\end{eqnarray}
The resulting explicit expressions for the twist-3 parameters are
rather bulky, so we do not give them explicitly.

\section{Sum Rules for Twist-2 Matrix Elements}\label{app:B}
\setcounter{equation}{0}

In this appendix we list and evaluate the QCD sum rules for twist-2
matrix elements of the $K^*$. 
The sum rules for $f_{K^*}^{\parallel,\perp}$, including 
SU(3)-breaking corrections, were
calculated in Refs.~\cite{BZ05,govaerts}, those for
$a_{1}^{\parallel,\perp}(K^*)$ in
Ref.~\cite{BZ05}, and those for $a_{2}^{\parallel,\perp}(K^*)$ 
in Ref.~\cite{BB03},
apart from the perturbative terms in $m_s^2$ and the 
radiative corrections to the quark condensate, which are new. 

For the
longitudinal parameters, the sum
rules read:
\begin{eqnarray}
\lefteqn{(f_{K^*}^\parallel)^2 e^{-m_{K^*}^2/M^2}  = 
\frac{1}{4\pi^2}\int\limits_{m_s^2}^{s_0}
ds\,e^{-s/M^2} \,\frac{(s-m_s^2)^2 (s+2m_s^2)}{s^3} +
\frac{\alpha_s}{\pi}\, \frac{M^2}{4\pi^2}\left( 1 -
e^{-s_0/M^2}\right)}\nonumber\\
&&{} +\frac{m_s\langle \bar s s\rangle}{M^2}\left(1 +
\frac{m_s^2}{3M^2} - 
\frac{13}{9}\,\frac{\alpha_s}{\pi}\right)+
\frac{4}{3}\,\frac{\alpha_s}{\pi} \, \frac{m_s\langle \bar q
  q\rangle}{M^2}+\frac{1}{12M^2}\,
\langle\frac{\alpha_s}{\pi}\,G^2\rangle 
\nonumber\\
&&{}  -\frac{16\pi\alpha_s}{9M^4}\,
\langle \bar q q\rangle\langle \bar s s\rangle +
\frac{16\pi\alpha_s}{81M^4}\,\left( \langle \bar q q\rangle^2 +
\langle \bar s s\rangle^2 \right),\\
\lefteqn{a_1^\parallel(K^*) (f_{K^*}^\parallel)^2 e^{-m_{K^*}^2/M^2} = 
\frac{5}{4\pi^2}\,m_s^4\int\limits_{m_s^2}^{s_0}
ds\,e^{-s/M^2} \,\frac{(s-m_s^2)^2}{s^4}}\nonumber\\
\hskip-10pt&&
{}+\frac{5m_s^2}{18M^4}\,\left\langle\frac{\alpha_s}{\pi}\,G^2\right\rangle 
\left( -\frac{1}{2}+ \gamma_E - {\rm
  Ei}\left(-\frac{s_0}{M^2}\right)+ \ln\,\frac{m_s^2}{M^2} + 
\frac{M^2}{s_0}\left( \frac{M^2}{s_0} -1\right) e^{-s_0/M^2}\right)\nonumber\\
&&{}
-\frac{5}{3}\,\frac{m_s\langle \bar s
  s\rangle}{M^2}\left\{1 + 
\frac{\alpha_s}{\pi}\left[ -\frac{124}{27} +
\frac{8}{9} \left(1-\gamma_E + \ln\,\frac{M^2}{\mu^2} +
\frac{M^2}{s_0}\,e^{-s_0/M^2} + {\rm Ei}\left(-\frac{s_0}{M^2}\right)
\right)\right]\right\}  \nonumber\\
&&{}-\frac{5}{3}\,\frac{m_s^3\squark}{M^4}+\frac{20}{27}\,
\frac{\alpha_s}{\pi} \, \frac{m_s\langle \bar q
 q\rangle}{M^2} + \frac{5}{9}\, \frac{m_s\langle\bar s \sigma g Gs\rangle}{M^4}
+\frac{80\pi\alpha_s}{81M^4}\,\left( \langle \bar q q\rangle^2 -
\langle \bar s s\rangle^2 \right),\label{eq:SRa1}\\
\lefteqn{
a_2^\parallel(K^*) (f_{K^*}^\parallel)^2 e^{-m_{K^*}^2/M^2} = }\nonumber\\
&&{}\frac{7}{4\pi^2}\,m_s^4\int\limits_{m_s^2}^{s_0}
ds\,e^{-s/M^2} \,\frac{(s-m_s^2)^2(2m_s^2-s)}{s^5} +
\frac{7}{72\pi^2}\,\frac{\alpha_s}{\pi} \,M^2 (1-e^{-s_0/M^2})
+\frac{7}{36M^2}\,\left\langle\frac{\alpha_s}{\pi}\,G^2\right\rangle 
\nonumber\\
&&{}+\frac{7}{3}\,\frac{m_s\langle \bar s
  s\rangle}{M^2}\left\{1+ 
\frac{\alpha_s}{\pi}\left[ -\frac{184}{27} +
\frac{25}{18} \left(1-\gamma_E + \ln\,\frac{M^2}{\mu^2} +
\frac{M^2}{s_0}\,e^{-s_0/M^2} + {\rm Ei}\left(-\frac{s_0}{M^2}\right)
\right)\right]\right\}  \nonumber\\
&&{}+\frac{49}{27}\,\frac{\alpha_s}{\pi} \, \frac{m_s\langle \bar q
 q\rangle}{M^2} - \frac{35}{18}\, 
\frac{m_s\langle\bar s \sigma g Gs\rangle}{M^4}
+\frac{224\pi\alpha_s}{81M^4}\,\left( \langle \bar q q\rangle^2 +
\langle \bar s s\rangle^2 \right) -
\frac{112\pi\alpha_s}{27M^4}\,\langle \bar q q\rangle \langle \bar s
s\rangle .\label{eq:SRa2}
\end{eqnarray}
For the transverse parameters, one has:
\begin{eqnarray}
\lefteqn{(f_{K^*}^\perp)^2 e^{-m_{K^*}^2/M^2}
= \frac{1}{4\pi^2}\int\limits_{m_s^2}^{s_0}
ds\,e^{-s/M^2} \,\frac{(s-m_s^2)^2 (s+2m_s^2)}{s^3} }\nonumber\\
&&{}+ \frac{1}{4\pi^2}\int\limits_{0}^{s_0}
ds\,e^{-s/M^2} \,\frac{\alpha_s}{\pi}\left( \frac{7}{9} +
\frac{2}{3}\,\ln \frac{s}{\mu^2}\right) 
-\frac{1}{12M^2}\,\langle\frac{\alpha_s}{\pi}\,G^2\rangle \nonumber\\
&&{} +\frac{m_s\langle \bar s
  s\rangle}{M^2}\left\{1+\frac{m_s^2}{3M^2}+
\frac{\alpha_s}{\pi}\left(-\frac{22}{9} + \frac{2}{3}
\left[ 1-\gamma_E + \ln\,\frac{M^2}{\mu^2} +
  \frac{M^2}{s_0}\,e^{-s_0/M^2} + {\rm Ei}\left(-\frac{s_0}{M^2}\right)\right]
\right)\right\}\nonumber\\
&&{}
-\frac{1}{3M^4}\,m_s\langle \bar s\sigma gGs\rangle -
\frac{32\pi\alpha_s}{81M^4}\,\left( \langle \bar q q\rangle^2 +
\langle \bar s s\rangle^2 \right),\label{eq:fKT}\\
\lefteqn{a_1^{\perp}(K^*)(f_{K^*}^\perp)^2 e^{-m_{K^*}^2/M^2} =  
\frac{5}{4\pi^2}\,m_s^4\int\limits_{m_s^2}^{s_0}
ds\,e^{-s/M^2} \,\frac{(s-m_s^2)^2}{s^4}+ \frac{10}{9}\, 
\frac{m_s\langle\bar s \sigma g Gs\rangle}{M^4}} \nonumber\\[-7pt]
&&{}+\frac{5m_s^2}{9M^4}\,\langle\frac{\alpha_s}{\pi}\,G^2\rangle 
\left( \frac{1}{4}+ \gamma_E - {\rm
  Ei}\left(-\frac{s_0}{M^2}\right)+ \ln\,\frac{\mu^2}{M^2} + 
\frac{M^2}{s_0}\left( \frac{M^2}{s_0} -1\right) e^{-s_0/M^2}\right)\nonumber\\
&&{}-\frac{5}{3}\,\frac{m_s\langle \bar s
  s\rangle}{M^2}\left\{1+ 
\frac{\alpha_s}{\pi}\left[ -\frac{49}{9} +
\frac{4}{3} \left(1-\gamma_E + \ln\,\frac{M^2}{\mu^2} +
\frac{M^2}{s_0}\,e^{-s_0/M^2} + {\rm Ei}\left(-\frac{s_0}{M^2}\right)
\right)\right]\right\},\label{eq:a1perp}\\
\lefteqn{
a_2^\perp(K^*) (f_{K^*}^\perp)^2 e^{-m_{K^*}^2/M^2} = }\nonumber\\
&&{}\frac{7}{4\pi^2}\,m_s^4\int\limits_{m_s^2}^{s_0}
ds\,e^{-s/M^2} \,\frac{(s-m_s^2)^2(2m_s^2-s)}{s^5} +
\frac{7}{90\pi^2}\,\frac{\alpha_s}{\pi} \,M^2 (1-e^{-s_0/M^2})
+\frac{7}{54M^2}\,\left\langle\frac{\alpha_s}{\pi}\,G^2\right\rangle 
\nonumber\\
&&{}+\frac{7}{3}\,\frac{m_s\langle \bar s
  s\rangle}{M^2}\left\{1+ 
\frac{\alpha_s}{\pi}\left[ -\frac{206}{27} +
\frac{16}{9} \left(1-\gamma_E + \ln\,\frac{M^2}{\mu^2} +
\frac{M^2}{s_0}\,e^{-s_0/M^2} + {\rm Ei}\left(-\frac{s_0}{M^2}\right)
\right)\right]\right\}  \nonumber\\
&& - \frac{49}{18}\, 
\frac{m_s\langle\bar s \sigma g Gs\rangle}{M^4}
+\frac{112\pi\alpha_s}{81M^4}\,\left( \langle \bar q q\rangle^2 +
\langle \bar s s\rangle^2 \right).
\end{eqnarray}
For $\rho$ and $\phi$, one has $a_1^{\parallel,\perp}=0$. To obtain
the sum rules for $f_{\phi}^{\parallel,\perp}$ and
$a_2^{\parallel,\perp}(\phi)$, one has to replace the perturbative
contributions to the above sum rules by
\begin{eqnarray}
\mbox{for $(f_{\phi}^{\parallel,\perp})^2$:~~}&&
 \frac{1}{4\pi^2}\int_{4m_s^2}^{s_0} ds \,e^{-s/M^2} \frac{(s+2 m_s^2)
 \sqrt{ 1-4 m_s^2/s}}{s}\,,\nonumber\\
\mbox{for $a_2^{\parallel,\perp}(\phi)(f_{\phi}^{\parallel,\perp})^2$:~~}&&
 -\frac{7}{2\pi^2}\int_{4m_s^2}^{s_0} ds \,e^{-s/M^2} \frac{m_s^4
 \sqrt{ 1-4 m_s^2/s}}{s^2}\,.
\end{eqnarray}
In addition, one has to substitute $\langle \bar qq\rangle\to \langle
\bar s s\rangle$ and to double the terms in $m_s\langle
\bar s s\rangle$, $m_s\langle \bar q q\rangle$ and $m_s \langle \bar s
\sigma g G s \rangle$.

We evaluate the sum rules using the input given in
Table~\ref{tab:QCDSRinput}.
The results are given in Sec.~\ref{sec:2.3} and Tab.~\ref{tab:num}.

\begin{table}[bt]
\renewcommand{\arraystretch}{1.3}
\addtolength{\arraycolsep}{3pt}
$$
\begin{array}{r@{\:=\:}l||r@{\:=\:}l}
\hline 
\quark & (-0.24\pm0.01)^3\,\mbox{GeV}^3 & \squark & (1-\delta_3)\,\quark\\
\mixed & m_0^2\,\quark &  \smixed & (1-\delta_5)\mixed\\[6pt]
\displaystyle \gluon & (0.012\pm 0.003)\, 
{\rm GeV}^4 & \multicolumn{2}{l}{}\\[6pt]\hline
\multicolumn{4}{c}{m_0^2 = (0.8\pm 0.1)\,{\rm GeV}^2,\quad \delta_3
  = 0.2\pm 0.2, \quad \delta_5 = 0.2\pm 0.2}\\\hline
\multicolumn{4}{c}{\overline{m}_s(2\,\mbox{GeV}) = (100\pm
20)\,\mbox{MeV}~~~\longleftrightarrow~~~ \overline{m}_s(1\,\mbox{GeV})
= (133\pm 27)\,\mbox{MeV}}\\
\multicolumn{4}{c}{\overline{m}_q(\mu) = \overline{m}_s(\mu)/R,
  \quad R = 24.6\pm 1.2}\\\hline
\multicolumn{4}{c}{\alpha_s(m_Z) = 0.1176\pm 0.002  ~\longleftrightarrow~ 
\alpha_s(1\,\mbox{GeV}) = 0.497\pm 0.005}\\\hline
\end{array}
$$
\renewcommand{\arraystretch}{1}
\addtolength{\arraycolsep}{-3pt}
\vskip-10pt
\caption[]{\small Input parameters for sum rules at the
  renormalisation scale $\mu=1\,$GeV. The value of $m_s$ is obtained
  from 
  unquenched lattice calculations with $n_f=2$ flavours 
as summarised in \cite{mslatt}, which agrees with the results from QCD
  sum rule calculations \cite{jamin}. $\overline{m}_q$ is taken from
  chiral perturbation theory \cite{chPT}.
$\alpha_s(m_Z)$ is the PDG
  average \cite{PDG}.}\label{tab:QCDSRinput}
\end{table}

\section{Sum Rules for Twist-3 Matrix Elements}\label{app:C}
\setcounter{equation}{0}

The chiral-even twist-3 parameters $\zeta_{3K^*}^\parallel$,
$\widetilde\omega_{3K^*}^\parallel$, $\widetilde\lambda_{3K^*}^\parallel$
can be determined from the correlation function
\begin{equation}\label{C.1}
i g_{\alpha\mu}^{\perp} \int d^4y \,e^{-ipy} \langle 0 | T \bar q(z) g
\widetilde G_{\alpha z} (vz) \gamma_z \gamma_5 s(0) \bar s(y)
\gamma_\mu q(y) | 0 \rangle = (pz)^2 (2-D)
\widetilde\Pi^\parallel_{3;K^*}(v,pz)\,;
\end{equation}
$D$ is the number of dimensions. 
In terms of hadronic contributions, the correlation function is given by
\begin{equation}
\widetilde\Pi_{3;K^*}^\parallel(v,pz) =
\frac{(f_{K^*}^\parallel)^2
  m_{K^*}^2}{m_{K^*}^2-p^2} \int{\cal
  D}(\underline{\alpha})\,e^{-ipy(\alpha_2+v\alpha_3)}\,
\widetilde\Phi_{3;K^*}^\parallel(\underline{\alpha}) + \dots;
\end{equation}
the dots denote contributions from higher-mass states.
The parameters $\kappa_{3K^*}^\parallel$, $\omega_{3K^*}^\parallel$ and
$\lambda_{3K^*}^\parallel$ can be obtained from an analogous correlation
functions $\Pi_{3;K^*}^\parallel(v,pz)$ with
$$
g\widetilde G_{\alpha z} \gamma_z\gamma_5 \to g G_{\alpha z}
i \gamma_z\,.
$$
For the chiral-odd operator, one has
\begin{equation}
i\int d^4y e^{-ipy} \langle 0 | T \bar q(z) \sigma_{z\mu} g
G_{z\mu}(vz) s(0) \bar s(y) \sigma_{pz} q(y) | 0 \rangle
= i (pz)^3 \Pi^\perp_{3;K^*}(v,pz)\,. 
\label{C.3}
\end{equation}
All three correlation functions $\Pi$ can be written as
$$\Pi_{3;K^*}(v,pz) = \int {\cal D}\underline{\alpha}
e^{-i pz (\alpha_2 + v \alpha_3)}\pi_{3;K^*}(\underline{\alpha})\,.
$$
For the functions $\pi_{3;K^*}(\underline{\alpha})$ we find
\begin{eqnarray}
\pi^{\perp }_{3;K^{*}}\left(\underline{\alpha}\right)
&=&
\frac{\alpha_{s}}{2\pi^{3}}\ln\frac{-p^2}{\mu^{2}}
\left[p^2\alpha_{1}\alpha_{2}\alpha_{3}\left(\frac{1}{\bar{\alpha}_{2}}-
\frac{1}{\bar{\alpha}_{1}}\right)\right.
\nonumber \\
&+&m_{s}m_{q}\frac{\alpha_{3}^{2}}{\bar{\alpha}_{1}\bar{\alpha}_{2}}
\left[\bar{\alpha}_{2}\left(\ln\frac{\alpha_{2}\alpha_{3}}{\bar{\alpha}_{1}}+
\frac{1}{2}\ln\frac{-p^2}{\mu^{2}}\right)-\left\{\alpha_{1}
\leftrightarrow\alpha_{2}\right\}\right]
\nonumber \\
&+&m_{s}^{2}\left\{-\alpha_{2}\alpha_{3}\left(\frac{1}{\bar{\alpha}_{2}}-
\frac{1}{\bar{\alpha}_{1}}\right)-\frac{\alpha_{2}\alpha_{3}^{2}}{
\bar{\alpha}_{2}^{2}}\left(\ln\frac{\alpha_{1}\alpha_{3}}{\bar{\alpha}_{2}}+
\frac{1}{2}\ln\frac{-p^2}{\mu^{2}}\right)\right\}
\nonumber \\
&-&m_{q}^{2}\left\{\alpha_{1}\leftrightarrow\alpha_{2}\right\}] 
\nonumber \\
&+&\frac{1}{12}\langle\frac{\alpha_{s}}{\pi}G^{2}\rangle
\frac{\alpha_{1}\alpha_{2}\left(\alpha_{1}-\alpha_{2}\right)\delta
\left(\alpha_{3}\right)}{\alpha_{1}m_{q}^{2}+\alpha_{2}m_{s}^{2}-
\alpha_{1}\alpha_{2}p^2}
\nonumber \\
&+&\frac{2}{3p^2}\frac{\alpha_{s}}{\pi}\left\{\right.
\frac{\bar{\alpha}_{3}}{2}\left(1+\alpha_{3}\right)\left(m_{q}
\langle\bar{q}q\rangle\delta\!\left(\alpha_{2}\right)-m_{s}\langle\bar{s}s
\rangle\delta\!\left(\alpha_{1}\right)\right) 
\nonumber \\
&+&\alpha_{3}\left[1+\alpha_{3}\left(1+\ln\left(\alpha_{3}
\bar{\alpha}_{3}\right)+\ln\frac{-p^2}{\mu^{2}}\right)\right]\left(m_{s}
\langle\bar{q}q\rangle\delta\!\left(\alpha_{2}\right)-m_{q}\langle\bar{s}s
\rangle\delta\!\left(\alpha_{1}\right)\right)\left.\right\} 
\nonumber \\
&+&\frac{1}{6
  p^4}\delta\!\left(\alpha_{3}\right)\left\{m_{q}\langle\bar{q}
\sigma g G q\rangle\delta\!\left(\alpha_{2}\right)-m_{s}\langle\bar{s}
\sigma g G s\rangle\delta\!\left(\alpha_{1}\right)\right\}
\nonumber \\
&+&\frac{16}{27p^4} \pi \alpha_{s} \delta\!\left( \alpha_{3}\right) 
\left\{\langle\bar{q}
q\rangle^{2}\delta\!\left(\alpha_{2}\right)-\langle\bar{s}s\rangle^{2}
\delta\!\left(\alpha_{1}\right)\right\},
\\
\pi^{\parallel}_{3;K^{*}}\left(\underline{\alpha}\right)
&=&
\frac{\alpha_{s}}{4\pi^{3}}\ln\frac{-p^2}{\mu^{2}}\left[\right.p^2
\alpha_{1}\alpha_{2}\alpha_{3}\left(\frac{1}{\bar{\alpha}_{2}}-
\frac{1}{\bar{\alpha}_{1}}\right)
\nonumber \\
&+&m_{s}m_{q}\frac{\alpha_{3}^{2}}{\bar{\alpha}_{1}\bar{\alpha}_{2}}
\left\{\bar{\alpha}_{2}\left(\ln\frac{\alpha_{2}\alpha_{3}}{\bar{\alpha}_{1}}+
\frac{1}{2}\ln\frac{-p^2}{\mu^{2}}\right)-\left\{\alpha_{1}
\leftrightarrow\alpha_{2}\right\}\right\}
\nonumber \\
&+&m_{s}^{2}\left\{-\alpha_{2}\alpha_{3}\left(\frac{1}{\bar{\alpha}_{2}}-
\frac{1}{\bar{\alpha}_{1}}\right)-\frac{\alpha_{2}\alpha_{3}^{2}}{
\bar{\alpha}_{2}^{2}}\left(\ln\frac{\alpha_{1}\alpha_{3}}{\bar{\alpha}_{2}}+
\frac{1}{2}\ln\frac{-p^2}{\mu^{2}}\right)\right\}
\nonumber \\
&-&m_{q}^{2}\left\{\alpha_{1}\leftrightarrow\alpha_{2}\right\}\left.\right]
\nonumber \\
&+&\frac{1}{24}\langle\frac{\alpha_{s}}{\pi}G^{2}\rangle\frac{\alpha_{1}
\alpha_{2}\left(\alpha_{1}-\alpha_{2}\right)\delta\!\left(\alpha_{3}\right)}{
\alpha_{2}m_{s}^{2}+\alpha_{1}m_{q}^{2}-\alpha_{1}\alpha_{2}p^2}
\nonumber\\
&+&\frac{1}{3p^2}\frac{\alpha_{s}}{\pi}\left\{\right.\frac{
\bar{\alpha}_{3}}{2}\left(1+\alpha_{3}\right)\left(m_{q}\langle\bar{q}q
\rangle\delta\!\left(\alpha_{2}\right)-m_{s}\langle\bar{s}s\rangle\delta
\left(\alpha_{1}\right)\right) 
\nonumber \\
&+&\alpha_{3}\left[1+\alpha_{3}\left(\ln\left(\alpha_{3}\bar{\alpha}_{3}
\right)+\ln\frac{-p^2}{\mu^{2}}\right)\right]\left(m_{s}\langle\bar{q}q
\rangle\delta\!\left(\alpha_{2}\right)-m_{q}\langle\bar{s}s\rangle\delta
\left(\alpha_{1}\right)\right)\left.\right\}
\nonumber \\
&+&\frac{1}{12 p^4}\delta\!\left(\alpha_{3}\right)\left\{
m_{q}\langle\bar{q}
\sigma g Gq\rangle\delta\!\left(\alpha_{2}\right)-m_{s}\langle\bar{s}
\sigma g G s\rangle\delta\!\left(\alpha_{1}\right)\right\}
\nonumber \\
&+&\frac{8}{27 p^4}\alpha_{s}\pi \delta\!\left( \alpha_{3}\right)
\left(\langle\bar{q}q\rangle^{2}\delta\!\left(\alpha_{2}\right) -
\langle\bar{s}s\rangle^{2}\delta\!\left(\alpha_{1}\right) \right),
\\
\widetilde{\pi}^{\parallel}_{3;K^{*}}\left(\underline{\alpha}\right)
&=&\frac{\alpha_{s}}{4\pi^{3}}\ln\frac{-p^2}{\mu^{2}}\left[\right.-
p^2\alpha_{1}\alpha_{2}\alpha_{3}\left(\frac{1}{\bar{\alpha}_{1}}+
\frac{1}{\bar{\alpha}_{2}}\right)
\nonumber \\
&+&m_{s}m_{q}\frac{\alpha_{3}^{2}}{\bar{\alpha}_{1}\bar{\alpha}_{2}}
\left\{\bar{\alpha}_{1}\left(\ln\frac{\alpha_{1}\alpha_{3}}{\bar{\alpha}_{2}}-
\frac{1}{2}\ln\frac{-p^2}{\mu^{2}}\right)+\left\{\alpha_{1}
\leftrightarrow\alpha_{2}\right\}\right\}
\nonumber \\
&+&m_{s}^{2}\left\{\alpha_{2}\alpha_{3}\left(\frac{1}{\bar{\alpha}_{1}}+
\frac{1}{\bar{\alpha}_{2}}\right)+\frac{\alpha_{2}\alpha_{3}^{2}}{
\bar{\alpha}_{2}^{2}}\left(\ln\frac{\alpha_{1}\alpha_{3}}{\bar{\alpha}_{2}}+
\frac{1}{2}\ln\frac{-p^2}{\mu^{2}}\right)\right\}
\nonumber \\
&+&m_{q}^{2}\left\{\alpha_{1}\leftrightarrow\alpha_{2}\right\}\left.\right]
\nonumber \\
&+&\frac{1}{24}\langle\frac{\alpha_{s}}{\pi}G^{2}\rangle\frac{\alpha_{1}
\alpha_{2}\delta\!\left(\alpha_{3}\right)}{\alpha_{2}m_{s}^{2}+\alpha_{1}
m_{q}^{2}-\alpha_{1}\alpha_{2}p^2}
\nonumber\\
&+&\frac{1}{3p^2}\frac{\alpha_{s}}{\pi}\left\{\right.\frac{
\bar{\alpha}_{3}^{2}}{2}\left(m_{s}\langle\bar{s}s\rangle\delta\!\left(
\alpha_{1}\right)+m_{q}\langle\bar{q}q\rangle\delta\!\left(\alpha_{2}\right)
\right) 
\nonumber \\
&+&\alpha_{3}\left[1-\alpha_{3}\left(2+\ln\left(\alpha_{3}\bar{\alpha}_{3}
\right)+\ln\frac{-p^2}{\mu^{2}}\right)\right]\left(m_{s}\langle\bar{q}q
\rangle\delta\!\left(\alpha_{2}\right)+m_{q}\langle\bar{s}s\rangle\delta
\left(\alpha_{1}\right)\right)\left.\right\}
\nonumber \\
&+&\frac{1}{12 p^4}\delta\!\left(\alpha_{3}\right)\left\{ m_{q}\langle
\bar{q}\sigma g
Gq\rangle\delta\!\left(\alpha_{2}\right)+m_{s}\langle\bar{s}
\sigma g G s\rangle\delta\!\left(\alpha_{1}\right)\right\}
\nonumber \\
&+&\frac{8}{27 p^4}\alpha_{s}\pi \delta\!\left( \alpha_{3}\right)
\left(\langle\bar{q}q\rangle^{2}\delta\!\left(\alpha_{2}\right)+
\langle\bar{s}s\rangle^{2}\delta\!\left(\alpha_{1}\right) \right)
\nonumber\\
&+&\frac{2}{3 p^4}\alpha_{s}\pi \delta\!\left( \alpha_{1}\right) 
\delta\!\left( \alpha_{2}\right)\langle\bar{q}q\rangle\langle\bar{s}s
\rangle.
\end{eqnarray}
Here we have dropped all terms that vanish upon Borelisation. The
above expressions include all terms to second order in the quark
masses. One comment is in order concerning the contribution of the
gluon condensate. Upon integration over $\alpha_i$, and subsequent
expansion in powers of the quark masses, this contribution contains
terms in $m_{s,q}^2 \ln (m_{s,q}^2/(-p^2))$, which are long-distance
effects and must not appear in the short-distance operator product
expansion of the correlation functions (\ref{C.1}) and (\ref{C.3}). 
As discussed in Ref.~\cite{logms}, the appearance of these 
logarithmic terms is due to the fact that the above expressions are
obtained using Wick's theorem to calculate the condensate
contributions, which implies that the condensates are normal-ordered: 
$\langle O
\rangle = \langle 0 |\!:\!O\!:\! |0\rangle$. Recasting the operator
product expansion in terms of
non-normal-ordered operators, all infrared sensitive terms can be
absorbed into the corresponding condensates. Indeed, using \cite{logms}
$$
\langle 0 | \bar s g G s | 0 \rangle = \langle 0| \!:\!\bar s g G
s\!:\!| 0 \rangle +
\frac{m_s}{2}\, \ln\,\frac{m_s^2}{\mu^2} \,\langle 0 |\! :\!
\frac{\alpha_s}{\pi}\, G^2\!:\!| 0\rangle,
$$
and the corresponding formula for $q$ quarks,
all terms in $\ln m_{q,s}^2$ can be absorbed into the mixed
quark-quark-gluon condensate and the resulting short-distance coefficients
can be expanded in powers of $m_{q,s}^2$. In calculating the sum rules, we
hence will use
$$
\ln\,\frac{-p^2}{m_{q,s}^2}\to \ln \frac{-p^2}{\mu^2}\,.
$$

The QCD
sum rules for the three hadronic parameters $\kappa_{3K^*}^\perp$,
$\omega_{3K^*}^\perp$ and $\lambda_{3K^*}^\perp$ describing the DA
$\Phi^\perp_{3;K^*}$, Eq.~(\ref{3.11}), to NLO in conformal spin then read:
\begin{eqnarray}
\left(f_{K^*}^\perp\right)^2 m_{K^*}^2 e^{-m_{K^*}^2/M^2}
      \kappa_{3K^*}^\perp
& = & \int_0^{s_0} e^{-s/M^2} \int {\cal D}\underline{\alpha}\,
      \frac{1}{\pi}\, {\rm Im}_{s}
      \pi^\perp_{3;K^*}(\underline{\alpha})\,,
\nonumber\\
\left(f_{K^*}^\perp\right)^2 m_{K^*}^2 e^{-m_{K^*}^2/M^2}
      \,\frac{1}{14}\,\omega_{3K^*}^\perp
& = & \int_0^{s_0} e^{-s/M^2} \int {\cal D}\underline{\alpha}\,
      (\alpha_1-\alpha_2)\frac{1}{\pi}\, {\rm Im}_{s}
      \pi_{3;K^*}^\perp(\underline{\alpha})\,,
\nonumber\\
\left(f_{K^*}^\perp\right)^2 m_{K^*}^2 e^{-m_{K^*}^2/M^2}
      \,\frac{3}{28}\,\lambda_{3K^*}^\perp
& = & \int_0^{s_0} e^{-s/M^2} \int {\cal D}\underline{\alpha}\,
      \left(\alpha_3-\frac{3}{7}\right)\frac{1}{\pi}\, {\rm Im}_{s}
      \pi_{3;K^*}^\perp(\underline{\alpha})\,.
\end{eqnarray}
The formulas for the other parameters are analogous.

We evaluate the above sum rules in the Borel window $M^2=1\,{\rm
  GeV}^2$ to $2.5\,{\rm GeV}^2$ and using the following values of
  the continuum threshold $s_0$:
\begin{eqnarray}
s_0^\parallel (\rho) &=& (1.3\pm 0.3)\,{\rm GeV}^2\,,\quad 
s_0^\parallel (K^*) = (1.3\pm 0.3)\,{\rm GeV}^2\,,\quad
s_0^\parallel (\phi) = (1.4\pm 0.3)\,{\rm GeV}^2\,,\quad \nonumber\\
s_0^\perp (\rho) &=& (1.5\pm 0.3)\,{\rm GeV}^2\,,\quad 
s_0^\perp (K^*) = (1.6\pm 0.3)\,{\rm GeV}^2\,,\quad
s_0^\perp (\phi) = (1.7\pm 0.3)\,{\rm GeV}^2\,.\hspace*{20pt}
\end{eqnarray}
Numerical results, including the uncertainties from the variation of
$M^2$, $s_0$ and the input parameters of Tab.~\ref{tab:QCDSRinput}, 
are given in Tab.~\ref{tab:num}.

\section{Loop Integrals}\label{app:D}
\setcounter{equation}{0}

For the interested reader, we collect the
loop integrals needed for calculating the correlation
functions in App.~\ref{app:C}. 

At one loop, one has (with $z^2=0$) \cite{BB03}:
\begin{eqnarray}
\int \left[d^L k\right]  e^{i f_k kz} \,\frac{(kz)^n}{(k^2)^a
  ((k-p)^2)^b} & = & (-1)^{a+b} \left(-p^2\right)^{D/2-a-b} (pz)^n
  \,\frac{\Gamma(a+b-D/2)}{\Gamma(a)\Gamma(b)}
\nonumber\\
&& \times\int_0^1 dw\,
  e^{i(1-w) f_k pz}\, w^{D/2-1-b} (1-w)^{D/2+n-1-a}\,,\label{E.1}
\end{eqnarray}
where the integration measure is defined as 
$d^D k = i/(4\pi)^2 \left[d^L k\right]$ and $f_k$ is an arbitrary
numerical factor.

One also needs the following integral:
\begin{eqnarray}
\lefteqn{\int \left[d^L l\right]  e^{i f_l lz} \,\frac{(lp)(lz)^j}{(l^2)^c
  ((l-k)^2)^d}}
\nonumber\\
& = & (-1)^{\frac{D-4}{2}} \left(k^2\right)^{D/2-c-d} (kp) (kz)^j
  \,\frac{\Gamma(c+d-D/2)}{\Gamma(c)\Gamma(d)}\int_0^1 du\,
  e^{i(1-u) f_l kz}\, u^{D/2-1-d} (1-u)^{D/2+j-c}
\nonumber\\
&&{}+(-1)^{\frac{D-4}{2}} \left(k^2\right)^{D/2+1-c-d} (pz)(kz)^{j-1}
  \,\frac{\Gamma(c+d-D/2-1)}{2\Gamma(c)\Gamma(d)}
\nonumber\\
&&{}\times\int_0^1 du\,
  e^{i(1-u) f_l kz}\, u^{D/2-d} (1-u)^{D/2-1+j-c}\left( j + i f_l
  (1-u) (kz)\right)\,.\label{E.2}
\end{eqnarray}
Two-loop integrals are obtained by combining the above one-loop
integrals. To obtain the correlation functions as integrals over
${\cal D}\underline{\alpha}$, one has to perform a variable
transformation from $(u,w)$ to $(\alpha_1,\alpha_2)$, where the
precise transformation relations are fixed by the ``canonical'' form
of the exponential: $\exp(-ipz \{\alpha_2 + v
(1-\alpha_1-\alpha_2)\})$. Note that (\ref{E.2}) contains an extra
factor $kz$ in the last line which comes from the Taylor expansion of
the exponential. This factor can be made disappear 
by partial integration of the final result and hence does not appear
in the explicit formulas  for the
correlation functions given in App.~\ref{app:C}.

\end{document}